# Understanding The Reversible Electrodeposition of Al in Low-Cost Room Temperature Molten Salts

Regina Garcia-Mendez,[a,b] Jingxu Zheng, [c,d],* David C. Bock,[e,f] Cherno Jaye,[g] Daniel A. Fischer,[g] Amy C. Marschilok,[e,f,h,i] Kenneth J. Takeuchi,[e,f,h,i] Esther S. Takeuchi,[f,h,i] and Lynden A. Archer [a,b,c,*]

a. Department of Chemical and Biomolecular Engineering, Cornell University; Ithaca, NY 14853, USA.
b. Cornell Energy Systems Institute, Cornell University; Ithaca, NY 14853, USA.
c. Department of Materials Science and Engineering, Cornell University; Ithaca, NY 14853, USA.
d. Department of Physics, Massachusetts Institute of Technology; Cambridge, MA 02129, USA.
e. Interdisciplinary Science Department, Brookhaven National Laboratory; Upton, NY 11973, USA.
f. Institute for Electrochemically Stored Energy, Stony Brook University; Stony Brook NY 11794, USA.
g. National Synchrotron Light Source II, Brookhaven National Laboratory; Upton NY 11973, USA.
h. Department of Chemistry, Stony Brook University; Stony Brook NY 11794, USA.
i. Department of Materials Science and Chemical Engineering, Stony Brook University; Stony Brook NY 11794, USA.

*Corresponding authors:* Lynden A. Archer and Jingxu Zheng

**Email address(es):** laa25@cornell.edu; jkzheng@mit.edu

Progress and Potential
Rechargeable batteries composed of earth-abundant and inexpensive metal anodes such as aluminum will have a crucial role in the widespread integration of renewable energy sources into the electric grid and electric vehicles, and are relevant due to their potential to significantly improve the energy density, safety, and cost compared to current state-of-art Lithium-ion batteries. A key barrier remains the high cost and corrosive characteristics of the ionic liquid electrolytes — the only electrolytes where Al anodes exhibit high-enough levels of reversibility to be of practical interest.

We report that a family of low-cost electrolytes intentionally designed enable highly reversible plating and stripping of metallic aluminum inside electrochemical cells at room temperature. We show further that by tuning the ratio of $[AlCl_4]^-$ and $[Al_2Cl_7]^-$ ions in solution, it is possible to sustain the high reversibility over thousands of charge-discharge cycles. The electrolyte molecular design approach presented here lays the foundation for low melting point electrolytes with the desired Lewis acidity for highly reversible Al electrodeposition and dissolution in secondary batteries.

Summary
Aluminum is the most earth-abundant metal, is trivalent, is inert in ambient humid air, and has a density approximately four-times that of lithium at room temperature. These attributes together make it attractive as a candidate material for cost-effective, long-duration storage of electrical energy in batteries. Scientific discoveries in the past decade have established that secondary Al batteries can be created by paring an Al anode with a graphite or transition metal oxide cathode, in imidazolium-based, room-temperature ionic-liquid-$AlCl_3$ (IL) electrolytes. Here we report findings from a systematic study that sheds light on the structural requirements, physicochemical, and transport properties of the IL electrolytes responsible for the high reversibility of Al battery anodes. Significantly, we find that the most important interfacial and transport properties of these electrolytes can be achieved in other electrolytes, including ammonium-based molten salts that are available at costs as much as twenty-times lower than the IL-$AlCl_3$ melt. High Al reversibility in ammonium- and imidazolium-based electrolytes is specifically shown to require a critical ratio of the solvation species ($AlCl_4^-$ and $Al_2Cl_7^-$), where Lewis's acidity and beneficial interfacial reactions continuously etch the $Al_2O_3$ resistive interfacial layer and form beneficial solid electrolyte interphase at the anode. The findings yield a quantitative molecular-level understanding of Al electrodeposition in room-temperature ammonium-based molten salts. We report further that successful development of new electrolyte families that support high Al anode reversibility also provides a good platform for detailed studies of the working mechanisms of an intercalation graphite cathode using X-ray absorption spectroscopy. Our findings therefore open new opportunities for developing simple, cost-effective, room-temperature Al batteries that enable long-duration electrical energy storage.

Introduction

Electrochemical cells based on aluminum have been pursued for more than a decade as a promising technology for storing electrical energy at low cost . (*1*, *2*) The rationale for this interest is as straightforward as it is manifold. The low-cost and simplicity of the battery anode and of the most often used cathode materials (Al foil and a graphitic carbon sheet), mature manufacturing and recycling of Al, air stability of both anode and cathode, high volumetric energy density and Earth crust abundance of Al (13 kWh/L for Al versus 6, 4 and 3 kWh/L for Li, Zn and Na, respectively, and 8% abundance for Al *vs.* 0.0065, 0.0075, 2.3% for Li, Zn, and Na, respectively) are the leading attributes that differentiate Al batteries from other candidates of contemporary interest. Notwithstanding these beneficial features, Al-based batteries have historically failed to live up to the promise of the chemistry for mainly two reasons. First, the high bandgap (7 eV) $Al_2O_3$ coating, which forms spontaneously on Al, which protects it from chemical attack by atmospheric agents, passivates the metal in conventional liquid electrolytes. The result is that reversible plating/stripping of Al during charge and discharge of a secondary/rechargeable Al battery is limited to a small number of specialized, expensive electrolytes. Second, the redox active species in these electrolytes are known to be bulky four- ($[AlX_4]^-$) and seven- ($[Al_2X_7]^-$) fold coordinated Al species, which rules out most insertion type materials as cathode candidates because reversible de/insertion of the multivalent coordinated Al ions is difficult due to their low mobility. Here, X is most commonly a halogen.

Despite these challenges, a number of studies conducted particularly in the last decade(*1–3*) have demonstrated that it is possible to create rechargeable Al batteries in 1-ethyl-3-methylimidazolium (EMIM) - $AlCl_3$ ionic liquid (IL) electrolyte melts. Why these electrolytes are successful in enabling Al reversibility largely remains an open question. Lewis acidity, high ion mobility at the Al/electrolyte interface, and a melting point below room temperature are considered required properties of ionic liquid electrolytes. Initial analyses of the interphases formed on an Al electrode suggested, but not confirmed, that the first of these properties is important for etching away the passivating $Al_2O_3$ coating and enabling fast interfacial ion transport at the Al/electrolyte interface. One study reported that exposure of Al to an imidazolium-based IL electrolyte creates an ionically conducting interphase on the Al electrode that remains intact when the electrode is immersed in other electrolyte media, including aqueous electrolytes up to 50 cycles, however, the composition and transport properties of the ionically conducting interphase were not investigated.(*4*)

In considering how one might design lower-cost analogs of the IL electrolytes, we first note that the desirable IL electrolyte properties are not unique to the [EMIM] cation. The Lewis acidity is in fact controlled by the molar ratio between the IL salt ($[C]^+X^-$) and $AlCl_3$ (*e.g.*, $[C]^+Cl^- + AlCl_3 \rightarrow [C]^+[AlCl_4]^-$). This, in turn, sets the relative concentration of electrochemically active $[AlX_4]^-$ and $[Al_2X_7]^-$ species present at room temperature.(*5*) The low melting point of the $AlCl_3$-EMIMCl electrolyte melt in contrast, is conventionally attributed to the large size of the imidazolium cation in comparison with the Cl- anion; the size mismatch destabilizes the ion packing in the solid-state and therefore results in a lower melting point. The key discovery that motivates the present report is that the electrolyte properties required for high Al anode reversibility in electrochemical cells are independent of the specific chemical structure of the cation species $[C]^+$. This surprising finding opens new approaches for designing chemically simpler and thus, inexpensive electrolytes that enable highly reversible cycling of Al batteries.

For example, Angell *et al.* reported that mixtures of urea, *i.e.*, $CO(NH_2)_2$, or its derivatives and $AlCl_3$ produce ionic-liquid-analog electrolytes, which allow both Al plating/stripping at the anode and intercalation at some cathodes. The authors noted, however, that more pronounced and undesirable parasitic side reactions are observed in these electrolytes, (*6*, *7*) attributable perhaps to the higher chemical reactivity of the carbonyl group in urea,(*8*, *9*) in comparison with the EMIM cation containing only aromatic, imidazole ring, C-C and C-H bonds, which are in general deemed to be chemically stable.

We hypothesize that simple, low-cost electrolytes designed using ammonium-based salts of broken symmetry (*i.e.*, in which the four moieties connected to the nitrogen atom are not equivalent), can be designed to simultaneously exhibit melting points below room temperature and high solvation power for $AlCl_3$ and its analogs. Such electrolytes would therefore enable plating and stripping of Al inside electrochemical cells with comparable levels of reversibility as the IL melts investigated in earlier works.

Results and Discussion

As a first step to designing low-cost electrolytes, we consider ILs in which $[C]^+$ is a quaternary ammonium species; a variety of quaternary ammonium cations composed of short-chain alkyl groups, *e.g.*, methyl-, and ethyl- that are commercially available and at relatively moderate costs (<$100/kg). Manipulating the symmetry of the cation offers a powerful route towards achieving low melting point electrolytes with the desired Lewis acidity. The simplest symmetry breaking of the nitrogen atom is achievable by replacing one alkyl group with a hydrogen atom, making it a ternary amine. The hydrogen atom breaks the perfect tetrahedron that facilitates packing and alters the symmetry of the cation from a high-order $T_d$ point group to a low-order $C_{3v}$ point group. Even this modest change reduces the number of symmetry elements from 24 to 6. We note here that symmetry breaking (*10*, *11*) interferes with molecular packing and is an already studied approach for depressing the melting point of molecular crystals, a.k.a. *Carnelley's rule*.

To evaluate this hypothesis, we studied physical and electrochemical properties of Al in electrolytes based on tetramethylammonium chloride ($AlCl_3$-TetraMACl), Triethylamine-hydrochloride ($AlCl_3$-TriEAHCl), and Trimethylamine-hydrochloride ($AlCl_3$-TriMAHCl) melts, in which the symmetry of the ammonium ion is broken to progressively greater extents. As reference, we compare the studied properties to those measured using $AlCl_3$-EMIMCl, the most frequently used IL electrolyte in Al electrochemical cells. As illustrated in Fig. 1A, the measured melting points of the symmetry-broken quaternary ammonium species drop markedly with reduced symmetry of the cation; the values achieved for TriMAHCl are in fact comparable to those for EMIMCl. Significantly, we also find that Al electrodes are reversible in all of the studied electrolytes and that in Al plating/stripping experiments, the quaternary ammonium chloride-based electrolytes display high levels of reversibility (Al plating/stripping coulombic efficiencies ≥ 99.3% for 1000 cycles at practical areal capacities and current densities (1 mAh·cm$^{-2}$, 4 mA·cm$^{-2}$), which are comparable to those observed in the $AlCl_3$-EMIMCl electrolyte.

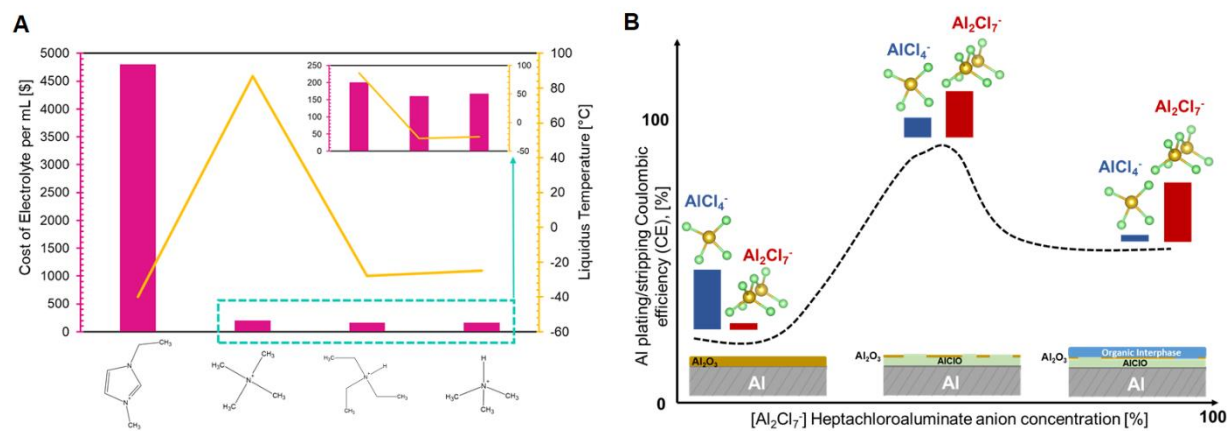

**Figure 1.** (**A**) Effect of molecular structure and symmetry on cost and melting point of electrolytes that enable reversible stripping and plating of Aluminum* (**B**) Schematic illustrating the interplay between the $AlCl_4^-$ to $Al_2Cl_7^-$ ratio and the Al plating/stripping coulombic efficiency coupled with the interphase chemistry on an Al electrode. *The concentration used for liquidus temperatures: 1.5 moles of $AlCl_3$ to 1 mole of alkylammonium/imidazolium chloride. The costs only consider the solvent and not the salt: $AlCl_3$.

Using $^{27}Al$ quantitative nuclear magnetic resonance (NMR), electrochemical measurements, and scanning probe microscopy, we show that plating/stripping reversibility increases with $AlCl_3$ concentration with an upper limit. By probing the surface chemistry of the anodes and imaging the electrode's morphology after electrodeposition using focused ion beam scanning electron microscopy (FIB-SEM), we find that Al plating/stripping reversibility is increased by continuous etching of the $Al_2O_3$ resistive interfacial layer and formation of a stable conductive solid electrolyte interphase (SEI) on the Al anode. The magnitude of the increase depended on the $AlCl_4^-$: $Al_2Cl_7^-$ ratio (Fig. 1B), where a balance between Lewis's acidity and lack of excess corrosion of the cell components determine the critical ratio. Our results quantitatively link the concentration of solvation species in the electrolytes to Al plating/stripping reversibility. In addition, the composition and resistivity of the SEI was deconvoluted through X-ray photoelectron spectroscopy and electrochemical impedance measurements as a function of the number of cycles.

We mapped out the concentrations of ammonium-based electrolytes that form ILs (or molten salts at room temperature) using phase diagrams obtained from differential scanning calorimeter (DSC) with heating and cooling capabilities (Fig. 2). The effect of the cooling/heating rate on freezing points in $AlCl_3$-EMIMCl (1.5: 1 in molar ratio) was evaluated to select the rate that is less dependent on the kinetics of the measurement. As seen in fig. S1, rates between 3 – 6 °C·min$^{-1}$ result in a minor change in the freezing point of the mixtures. Thus, 3 °C·min$^{-1}$ was chosen to construct phase diagrams in all systems.

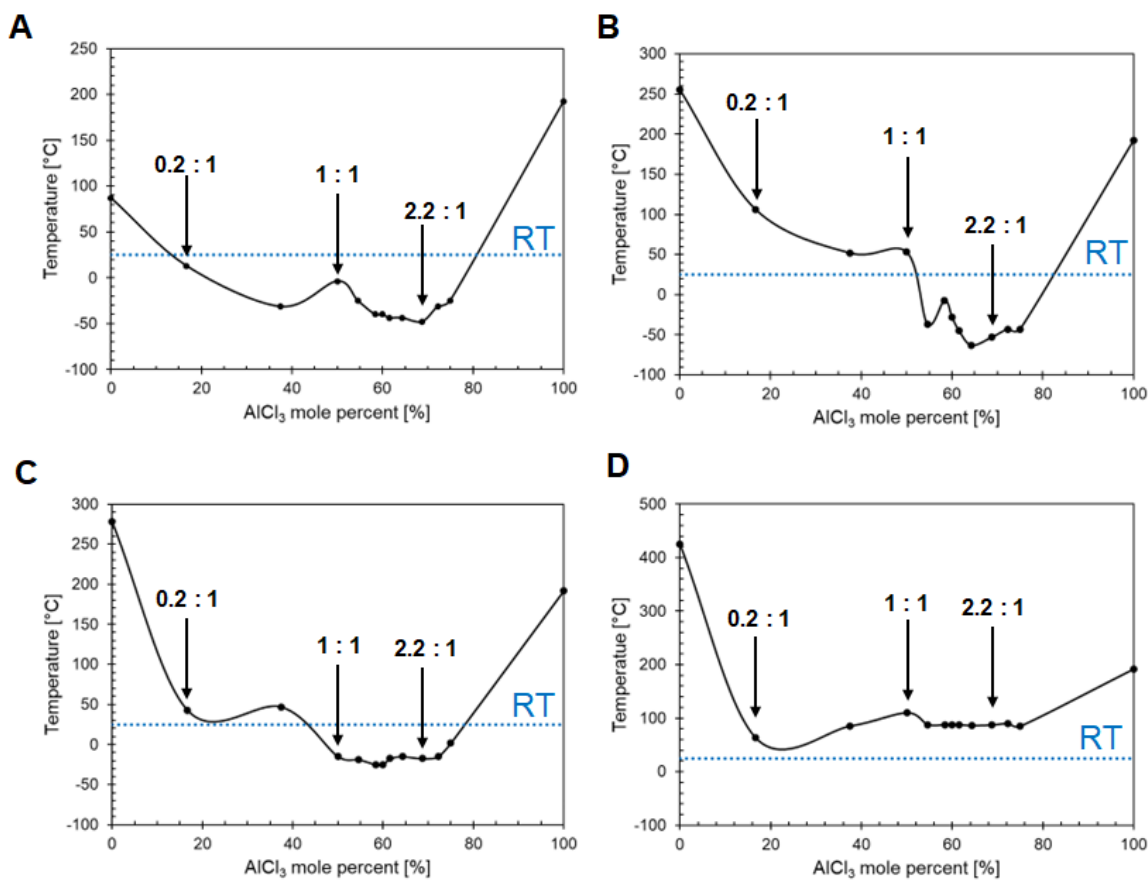

**Figure 2.** Phase diagrams of ammonium-based electrolytes as a function of AlCl$_3$ mole percent. (**A**) 1-ethyl-3-methylimidazolium chloride EMIMCl, (**B**) Triethylamine-hydrochloride (TriEAHCl), (**C**) Trimethylamine-hydrochloride (TriMAHCl), and (**D**) Tetramethylammonium chloride (TetraMACl). Blue dotted line corresponds to 25°C (RT = room temperature).

The concentrations that formed molten salts at room temperature were mainly observed between the AlCl$_3$ – salt ratios 1: 1 (50 mol % AlCl$_3$) to 2.6: 1 (72.2 mol % AlCl$_3$). The exception is the electrolyte system composed of the most symmetric and highest molar mass cation AlCl$_3$-TetraMACl, which remained in solid, crystalline form over the entire concentration range of AlCl$_3$ studied. These measurements are in agreement with our preliminary observations in mixing higher tetra-alkyl-based compounds with AlCl$_3$ (fig. S2), confirming the effect of molecular structure on the melting point and liquid ranges. (*12*) Increasing the molecular symmetry in the electrolyte, as is the case with tetra-alkyl compounds (fig. S3), promotes charge ordering and efficient packing remaining in crystalline form at room temperature.(*13*)

Using the group contribution method (GCM) developed by Lazzús, melting temperatures (T$_m$) for a wide range of ILs with cation groups including imidazolium, pyridinium, pyrrolidinium, ammonium, *etc*. can be estimated with an average deviation of 7%.(*13*) Thus, the GCM method was used to rationalize the structural effects on the thermal behavior of each system. The mathematical foundation of the GCM method is based on the principle of collinearity. T$_m$ is a linear combination of any number of functions of the independent variables or functional groups in this case. The functional groups are divided into groups for the cation and the anion part, and the size, shape, asymmetry of the cation and anion parts of each system play a role in the $T_m$ measured. The melting point decreased as the $AlCl_3$ concentration increased due to the formation of $Al_2Cl_7^-$ anions, which are larger than $AlCl_4^-$, known to be formed in acidic compositions, *i.e.,* when $AlCl_3$ is added in excess of the EMIMCl.(*14–16*) An increase in anion size led to reductions in the melting points by reducing the Coulombic attraction contributions to the lattice energy of the crystal and increasing the covalency of the ions.(*12*) For the cation part, length and asymmetry were more significant in determining the melting point and liquid ranges. The TriEAHCl systems having longer alkyl chains than the TriMAHCl enable higher charge delocalization, further reducing the overall charge density, and both of their asymmetries disrupts the crystal packing and reduces the crystal lattice energy resulting in a lower melting temperature. Given that the TetraMACl system remained crystalline for all concentration ranges, it was not further explored as an electrolyte for the room-temperature Al electrochemical cells of interest here.

We next interrogated the solvation species in the ammonium-based and imidazolium-based IL electrolytes using $^{27}Al$ NMR from chemical shifts; 103.8 ± 2.0 ppm for $AlCl_4^-$ (*6, 17–19*), and 97.5±1.0 ppm for $Al_2Cl_7^-$ (*18*) (figs S4-6). There was no detectable concentration of $Al_3Cl_{10}^-$ species, 81 ppm (*20*) throughout the entire $AlCl_3$ concentration range evaluated in this work. The spectra were fitted to quantitatively compare the relative concentration of the anionic solvation species for each sample that form molten salts at room temperature (Fig. 3). Both TriMAHCl and TriEAHCl systems exhibited the same aluminum anion species as the $AlCl_3$-EMIMCl system. Nevertheless, the ratio of each species in the electrolytes varies for each system, especially for the TriMAHCl system. Fig. 3A, B show that from the range at which the EMIMCl and TriEAHCl systems form a molten salt, there is a wide range of $AlCl_4^-$ and $Al_2Cl_7^-$ concentrations that can be attained by varying the $AlCl_3$ mole fraction. These observations are consistent with previous $^{27}Al$ NMR studies of $AlCl_3$-EMIMCl electrolytes.(*5, 21*) In contrast, Fig. 3C shows that the majority of the molten salts formed in the TriMAHCl system have $Al_2Cl_7^-$ as the majority species (≥ 64 %), *i.e.,* all acidic.

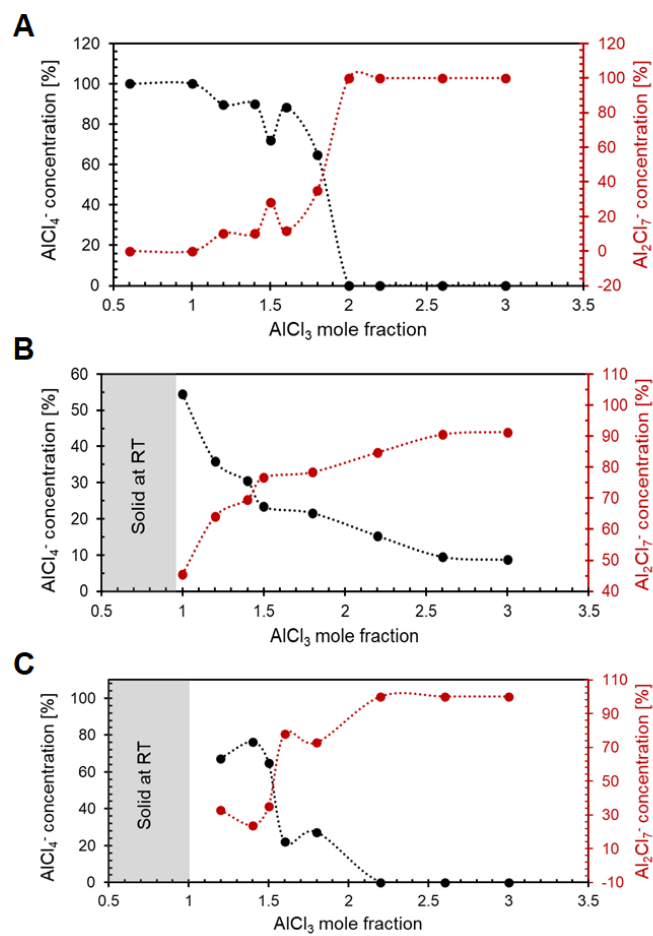

**Figure 3.** Percentage of $AlCl_4^-$ and $Al_2Cl_7^-$ in ionic liquids as a function of $AlCl_3$ mole fraction, determined via quantitative $^{27}Al$ NMR (**A**) $AlCl_3$-EMIMCl (1-ethyl-3-methyl-imidazolium chloride) (**B**) $AlCl_3$-TriEAHCl (Triethylamine hydrochloride) (**C**) $AlCl_3$-TriMAHCl (Trimethylamine hydrochloride). (See also Figures S4-6).

One of the critical challenges in developing rechargeable Al batteries is the limited number of electrolytes that enable reversible Al plating/stripping at room temperature.(*17*) We investigated the reversibility of Al in TriMAHCl and TriEAHCl electrolytes using galvanostatic experiments in which Al was deposited on a conductive fibrillar carbon substrate. The carbon 3D-electrode architecture was chosen since it has been demonstrated to facilitate electron transport and promote strong oxygen-mediated chemical bonding between Al deposits and the carbon substrate, facilitating control of deposition morphology and limiting out-of-plane Al growth in imidazolium-based electrolytes.(*3, 22*) Scanning electron microscopy (SEM) and energy dispersive spectroscopy (EDS) analysis of the substrate were conducted to study the chemistry of the electrodeposits. From the EDS elemental maps (fig. S7), it is evident that conformal coatings composed of Al is formed in the fibres of the carbon substrate for all systems. This finding is in agreement with the morphology observed in a prior study using 1.3: 1 $AlCl_3$-EMIMCl.(*3*)

The confirmation of Al electrodeposits in the proposed systems combined with the knowledge gained via NMR spectroscopy enables detailed investigation of the electrolyte features responsible for the observed reversibility, by simply varying the $AlCl_4^-$ and $Al_2Cl_7^-$ concentration. Prior studies reported that acidic $AlCl_3$-EMIMCl melts (*i.e.,* melts in which $AlCl_3$ is added in excess of the EMIMCl, and for which the dominant species is $Al_2Cl_7^-$) are responsible for the reversible Al deposition according to the reaction $4\ Al_2Cl_7^- + 3\ e^- \leftrightarrow Al + 7\ AlCl_4^-$.(*23*) Al plating/stripping Coulombic efficiency (reversibility) was measured by performing galvanostatic tests on carbon 3D-electrode substrates using selected compositions of each electrolyte system. The CE of each system quantifies the percentage of Al metal that can be stripped from the Al originally plated. In these studies, three concentrations were chosen for each electrolyte; one with $AlCl_4^-$ as the dominant species, one with the $Al_2Cl_7^-$ as the dominant species, and a third that consisted of a mixture of both anionic species. The electrolytes with a mixture of both anionic species, especially those with a higher concentration of $Al_2Cl_7^-$ (Fig. 4), showed higher reversibility for all electrolytes (fig. S8). It is noteworthy that the $AlCl_3$-EMIMCl mixture containing approximately 65% of $AlCl_4^-$ and 35% $Al_2Cl_7^-$, compared favorably to those, respectively, containing 22% and 23% of $AlCl_4^-$ and 78% and 77% of $Al_2Cl_7^-$ for the TriEAHCl and TriMAHCl mixtures. The Al plating/stripping coulombic efficiencies (CE) measured were 99.2%, 99.5% and 99.3%, for 1.8: 1 $AlCl_3$-EMIMCl, 1.6: 1 $AlCl_3$-TriEAHCl, and 1.5: 1 $AlCl_3$-TriMAHCl, respectively, at 1 mAh·cm$^{-2}$ areal capacity and 4 mA·cm$^{-2}$ current density for 1,000 cycles. Interestingly, we observed that these modest changes in CE are accompanied by quite large differences in cell polarization and morphology of the electrodeposited Al.The voltage profiles in Fig. 4B reveal more stable Al stripping/plating reaction for the TriEAHCl and TriMAHCl electrolytes by a reduced overpotential that remains constant from cycles ~50-1000, compared to the EMIMCl electrolyte. Comparable behavior was observed in Al symmetric cells (fig. S9). The increase in coulombic efficiency and narrowing of the overpotential over the first 50 cycles will be explained in detail in the next section via surface chemistry analyses via X-ray photoelectron spectroscopy (XPS).

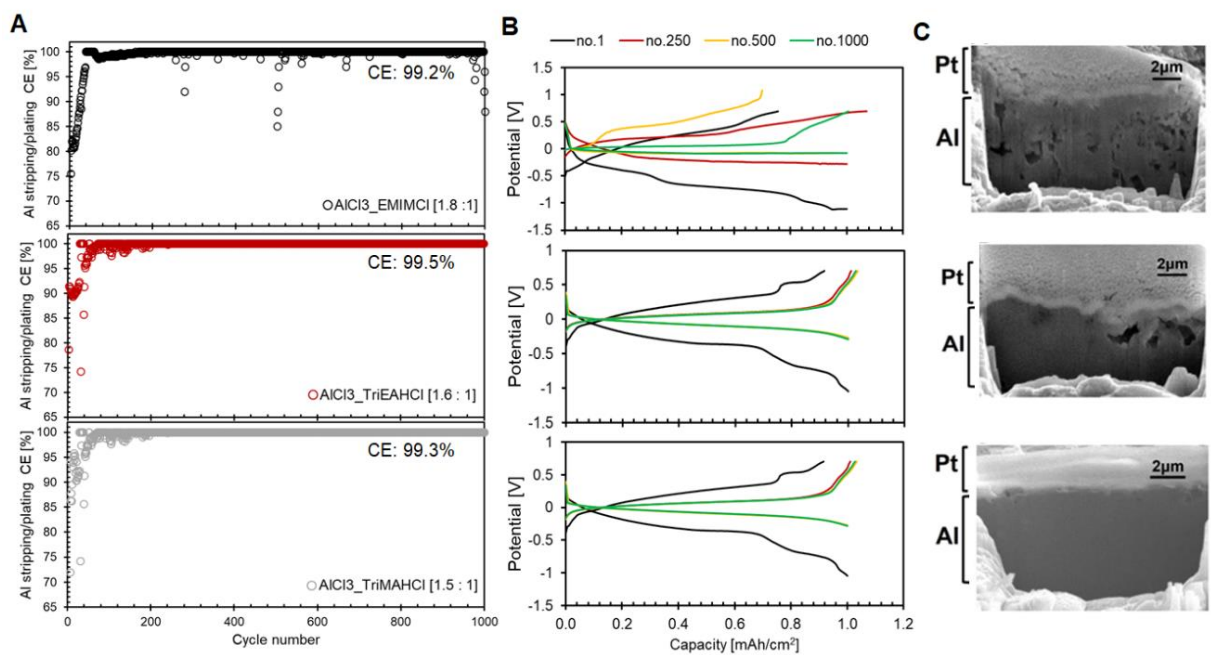

**Figure 4.** Electrochemical cycling behavior of Al electrodes in galvanostatic plating/stripping experiments using various electrolytes; (top) 1.8: 1 $AlCl_3$-EMIMCl (1-ethyl-3-methyl-imidazolium chloride) (middle) 1.6:1 $AlCl_3$-TriEAHCl (Triethylamine hydrochloride) (bottom) 1.5 : 1 $AlCl_3$-TriMAHCl (Trimethylamine hydrochloride) (**A**) Al plating /stripping efficiency measured at 1 mAh·cm$^{-2}$ areal capacity, 4 mA·cm$^{-2}$ current density (**B**) Voltage profiles obtained during Al plating/stripping at 1 mAh·cm$^{-2}$ areal capacity, 4 mA·cm$^{-2}$ current density (**C**) SEM of FIB-milled cross-sections of Al anodes cycled using various electrolytes. (See also Figures S7-10).

Representative SEM analysis on FIB-cut cross-sections of the electrodeposited Al confirmed the presence of voids near the surface and within the bulk of the anode for the EMIMCl electrolyte (Fig. 4C). Evidence of voids and pitting are apparent near the surface of the electrodeposited Al in the TriEAHCl. In contrast, electrodeposited Al harvested from the TriMAHCl electrolyte forms as a dense layer throughout the bulk and near the surface of the anode. Multiple areas in all electrolytes were examined to confirm that these observations are in fact quite general. Our results therefore show that ammonium-based electrolytes composed of a mixture of $AlCl_4^-$ and $Al_2Cl_7^-$, with $Al_2Cl_7^-$ as the moderate majority (≥ 70%) ionic species, facilitate higher reversibility of Al and formation of deposits with planar, dense morphologies. The mixtures with with the dominant species of either $AlCl_4^-$ or $Al_2Cl_7^-$ (> 65% of $AlCl_4^-$ or ≥ 90% of $Al_2Cl_7^-$) comparatively underperformed in Al plating/stripping reversibility (fig. S8). The mixtures with a predominantly higher concentration of $AlCl_4^-$ showed poor reversibility, with average CE values ranging from 64.9 to 75.7% (fig. S8 A, C).

This is a direct result of the inherent irreversibility of the predominant electrodeposition process in neutral melts: $AlCl_4^-$ + 3 e- ↔ Al + 4 Cl-. (*23*) As the $Al_2Cl_7^-$ concentration increases, the Lewis acidity of the electrolytes increases. Thus, a higher Lewis acidity corresponds to a higher affinity from the electrolytes to accept a pair of electrons and form a coordinate covalent bond, resulting in electrochemical corrosion of the metal parts in the cell. This phenomenon was observed in highly acidic electrolytes ($Al_2Cl_7^-$ ≥ 90%), where the charging step would extend for overly long periods, indicative of a faradaic process occurring, as seen in fig. S10. Although the CE values appeared to be higher than the mixtures with high $AlCl_4^-$ concentration, upon closer examination of the potential curves, electrochemical corrosion after tens of cycles was identified as the controlling process occurring within the cell.

Although the $AlCl_4^-$ ion cannot be reduced directly at the electrode to form Al, it can affect the Al electrodeposition process through a dissociation reaction: 2 $AlCl_4^-$ ↔ $Al_2Cl_7^-$ + Cl-. In turn, the reduction of $Al_2Cl_7^-$ leads to aluminum deposition according to the reaction: 4 $Al_2Cl_7^-$ + 3e- ↔ Al + 7 $AlCl_4^-$.(*17, 24*) A key remaining question concern the effect of the electrolyte composition on physicochemical and transport properties. The conductivity of a material is a measure of the available charge carriers and their mobility. Ionic liquids composed entirely of ions would be expected to have high ionic conductivity values. However, ion-pairing and/or ion aggregation commonly decrease the number of available charge carriers. (12) Prior work from Fannin *et al.* (*25*) reported ionic conductivity values at room temperature for the $AlCl_3$-EMIMCl system; the basic concentrations with a value of ~6.5 mS·cm$^{-1}$ (presence of Cl- and $AlCl_4^-$ anions), neutral with a value of ~23.0 mS·cm$^{-1}$ (presence of $AlCl_4^-$ anions), and acidic concentrations with a value of ~15.0 mS·cm$^{-1}$ (presence of $AlCl_4^-$ and $Al_2Cl_7^-$ anions). We measured the ionic conductivity of electrolytes used in the study using a dielectric spectrometer with a temperature controller incorporated. The values measured for the EMIMCl electrolytes followed the expected trend: the neutral concentration exhibited the highest conductivity, followed by acidic concentrations, and a lower conductivity for the only basic concentration considered in this work. Both TriMAHCl and TriEAHCl electrolyte mixtures with a high concentration of $AlCl_4^-$ presented the highest conductivity values and a decrease as the concentration of $Al_2Cl_7^-$ increased per $^{27}Al$ quant-NMR analysis. The TriMAHCl mixtures having $Al_2Cl_7^-$ as the majority species (≥ 64 %) exhibited lower conductivity values (4 – 21 mS·cm$^{-1}$) compared to the TriEAHCl mixtures (12 – 37 mS·cm$^{-1}$). Even though the large ion size from the triethylammonium cation would be expected to reduce ion mobility even further than a trimethylammonium cation, the decrease in the number of available charge carriers via the $AlCl_4^-$ dissociation reaction dominates.

The observed temperature-dependent conductivity behavior exhibited a classical linear Arrhenius behavior above room temperature (fig. S11), as observed in most ionic liquids. Generally, as the ionic liquids approach their glass transition temperatures, the conductivity displays a significant negative deviation from linear behavior, consistent with glass-forming liquids, best described using the empirical Vogel-Tammann-Fulcher (VTF) equation. (*26, 27*) In the temperature range evaluated, the linear behavior may be explained by postulating the onset of an ionic solid-like conductance mechanism. Thus, there may occur a process where fluctuations in configurational entropy determine the relaxation time, and a process where individual ions undergo successive displacement in a semirigid lattice. (*28*) Both processes contribute to the conductance and hence to the slope of the ln σ vs. 1000·T$^{-1}$ line, *i.e.*, the activation energy.

The activation energy for AlCl$_4^-$ and Al$_2$Cl$_7^-$ conduction through the AlCl$_3$-EMIMCl electrolytes varied based on the ions present in each mixture and their relative concentration. The electrolyte mixture with a 65% AlCl$_4^-$ and 35% Al$_2$Cl$_7^-$ showed the lowest activation energy (9.80 kJ·mol$^{-1}$) compared to the mixtures with AlCl$_4^-$ as the majority species or the Al$_2$Cl$_7^-$ as the majority species (14.35 and 11.64 kJ·mol$^{-1}$, respectively) in the AlCl$_3$-EMIMCl system. Interestingly, the 1.5: 1 AlCl$_3$-TriMAHCl electrolyte that exhibited a dense electrodeposited morphology compared to the rest presented a comparable activation energy for ionic conduction to the 1.6: 1 AlCl$_3$-TriEAHCl electrolyte and higher than the 1.8: 1 AlCl$_3$-EMIMCl electrolyte that exhibited a porous electrodeposited morphology. The activation energies calculated in combination with the ionic conductivities measured indicated that the high reversibility for plating and stripping is not a result of the ionic transport within the electrolyte but point out that it may be an interfacial-controlled phenomenon.

Progress in developing practical Al batteries has been hindered by multiple challenges associated with slow interfacial charge transport and sluggish electrochemical reactions at both the anode and the cathode. Among these challenges, the rapid formation of an irreversible, resistive, passivating Al$_2$O$_3$ film on the metal anode is considered the most difficult, because the oxidation reaction is thermodynamically favored, impedes stripping of Al$^{3+}$, and reduces the battery working voltage.(*23*, *29–31*) Imidazolium-based IL electrolytes have been reported to overcome the anode passivation problem.(*1*, *2*, *4*) These electrolytes are believed to etch the passivating layer on the Al anode surface and form a new interfacial layer (solid electrolyte interphase, SEI) that simultaneously protect Al and regulate transport of ions to the electrode. Thus, we interrogated the surface chemistry of the Al anodes after a hundred cycles of Al plating/stripping using SEM-EDS (fig. S12 and S13), followed by a more in-depth examination using X-Ray Photoelectron Spectroscopy (XPS).

EDS analyses showed that the oxygen content present in the imidazolium-based electrolytes is twice as high as the oxygen present in the ammonium-based electrolytes after 100 cycles of plating and stripping (ending with a plating step), suggesting that etching of the Al$_2$O$_3$ layer is favored in the ammonium-based electrolytes. The XPS high-resolution core scans provided insight into the nature of the bonds that each element detected participate in the surface chemistry evolution of the Al anode as the number of cycles increased, going from a lower CE until it stabilized at a higher value (Fig. 5 A-I).

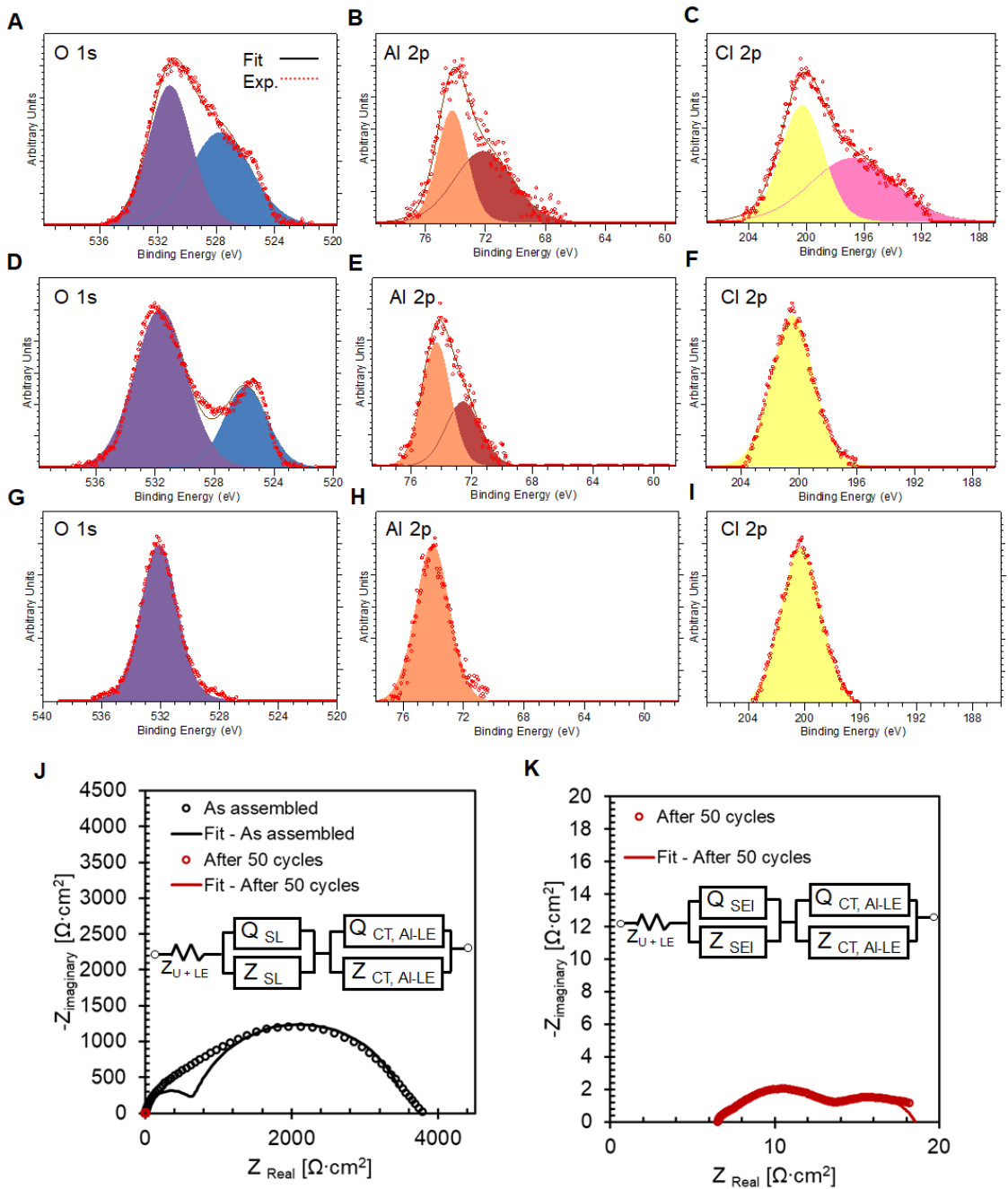

**Figure 5.** X-Ray Photoelectron Spectroscopy (XPS) analysis of Al anodes as a function of cycle number and depth on O 1s, Al 2p, and Cl 2p core levels and electrochemical impedance spectroscopy (EIS) measurements on Al symmetric cells (**A-C**) cycle no.5, (**D-F**) cycle no.50, (**G-I**) cycle no.50 after 127 min of Ar sputtering, (**J**) EIS spectra of Al symmetric cells as assembled (black trace) and after 50 cycles (red trace), (**K**) EIS spectrum of Al symmetric cell after 50 cycles of plating and stripping. (See also Figures S12-15).

The anodes cycled with the 1.5: 1 AlCl$_3$-TriMAHCl electrolyte were chosen for more in-depth studies since these anodes showed the more planar, dense electrodeposited morphology compared to the other systems.

In addition, Al foil as-received was analyzed as a baseline (fig. S14). High-resolution and depth profiling measurements were performed on Al anodes transferred without air exposure between an Ar-filled glovebox and the ultrahigh vacuum XPS chamber. The O 1s and Al 2p signals reveal a significant difference between the surface layer after galvanostatic cycling in $AlCl_3$-TriMAHCl electrolyte *vs.* as-received. The anode surface after 5 cycles of galvanostatic plating and stripping exhibited two chemical environments for both elements. The Cl 2 p signal was also evaluated due to the presence of chloroaluminate species detected *via* $^{27}$Al NMR in the electrolytes. To determine the nature of the bonds of each element, the Al-O-Cl phase diagram was evaluated since it determines the thermodynamically stable phases in this system. The binary/ternary stable phases according to their standard Gibbs free energy of formation energy are $Al_2O_3$ < AlClO < $AlCl_3$ < $ClO_2$ < $ClO_3$ < $Cl_2O_7$ ( -3.427, -2.785, -1.966, -0.353, -0.339, -0.303 eV· atom$^{-1}$, respectively). (*32*)

It is reasonable to expect that the phase that will form after etching the alumina phase, will be the one with the most negative standard Gibbs free energy of formation. In this case, AlClO, with an anticipated higher binding energy than the oxide given the Cl higher electronegativity. In addition, the atomic percent measured of each species, especially after > 2 h of Ar-sputtering (fig. S15) further confirmed that the higher binding energy bonding environment corresponds to AlClO and not $AlCl_3$, $Al(OH)_3$, AlO(OH) or $Cl_2$ species. The anode surface consists of predominantly AlClO species, while a greater concentration of $Al_2O_3$ exists on the as-received Al. This suggests that the ammonium-based electrolyte etches part of the $Al_2O_3$ layer and reacts with the Al anode upon galvanostatic cycling, forming a less resistive compound protecting the Al surface from further oxidation. Even though the resistance of the AlClO layer is not expected to be significantly less electronically resistive than the native $Al_2O_3$ given their band gaps (5.597 eV *vs.* 6.044 eV), (*32*) the ion conduction of chloroaluminate species through the AlClO layer is expected to be favored compared to an $Al_2O_3$ layer as it was observed through electrochemical impedance spectroscopy (EIS) measurements in Al symmetric cells (Fig. 5 J-K). As assembled, the cell exhibited two constant phase elements; one corresponding to the native $Al_2O_3$ layer, "SL" for solid layer (capacitance value of 1x10$^{-6}$ F, KHz range) and a second one corresponding to the charge transfer resistance between the Al anode and the liquid electrolyte, "Al-LE", (capacitance value of 12x10$^{-5}$ F, Hz - mHz range). The capacitance values obtained through the equivalent circuit modelling for each semi-circle are in good agreement with values for surface layers and sample-electrode interface transport phenomena. (*33*) Analogously, the EIS spectrum after 50 cycles of cycling (after the CE values have reached its equilibrium value and remain constant onwards), was analyzed using the same equivalent circuit model. However, in this case, the first semi-circle, is attributed to the newly formed AlClO layer ("SEI") confirmed by survey and high-resolution scans through XPS measurements on the Al anode, showing a significant lower impedance compared to the alumina layer with improved charge transfer impedance. Thus, the beneficial SEI layer in combination with the ease of charge exchange at the interface facilitates the kinetics of the electrodeposition process at the electrode-electrolyte interface, resulting in a more planar, dense electrodeposited morphology.

The anodes collected after 50 cycles of galvanostatic plating/stripping were analyzed before and after ~2 h of Ar-sputtering (~170 nm deep from the surface), (Fig. 5 D-F, and G-I). After 50 cycles, the CE reached its highest value and remained roughly constant. The amount of $Al_2O_3$ present at the surface was lower, while the AlClO increased, consistent with the signals observed in the O 1s, Al 2p and Cl 2p core levels. The closer to the bulk of the Al anode, the layer more closely resembled the presence of AlClO while the $Al_2O_3$ contribution disappeared. In total, these observations suggest that the Lewis' acidity of the electrolytes continuously etch the ionically resistive alumina layer, exposing the Al anode to the electrolyte, that in turn, reacts with the chloroaluminate species forming AlClO. However, there is an upper limit to the Lewis' acidity (*i.e.,* concentration of $Al_2Cl_7^-$ species) needed for this to occur. At high concentrations of $Al_2Cl_7^-$ (e.g., ≥ 90 %), corrosion of the components is observed (fig. S10) and readily electrochemical decomposition of the electrolyte is anticipated under reductive conditions (fig. S16) forming an organic interphase.

As a first demonstration of the practical relevance of our findings, Al *vs.* graphite full cells were evaluated in galvanostatic cycling experiments using the electrolytes that showed the highest Al plating/stripping reversibility: AlCl$_3$-TriMAHCl (1.5: 1) and AlCl$_3$-TriEAHCl (1.6 :1). We note that although the ratio of AlCl$_4^-$ to Al$_2$Cl$_7^-$ are similar in these electrolytes and the electrolytes exhibit comparable Al plating/stripping reversibility, only the AlCl$_3$-TriMAHCl (1.5: 1) supported stable long-term cycling of Al batteries, with more capacity fade when using AlCl$_3$-TriEAHCl (Fig 6A and figs S17-18). We conclude that the specific cation chemistry and physical properties play an important role in the solvation/de-solvation characteristics of the electrolytes. We will take this aspect up in a follow-up study in which the effects of TriMAH$^+$ and TriEAH$^+$ size on the de-solvation energy and formation of a cathode electrolyte interphase could provide insight into the cycling stability of both systems. In addition, the increase in reversibility in the first ~50 cycles is attributed to the anode interface, where the etching of the alumina surface layer occurs followed by the formation of a SEI consisting of AlClO that enables high reversibility of Al plating and stripping.

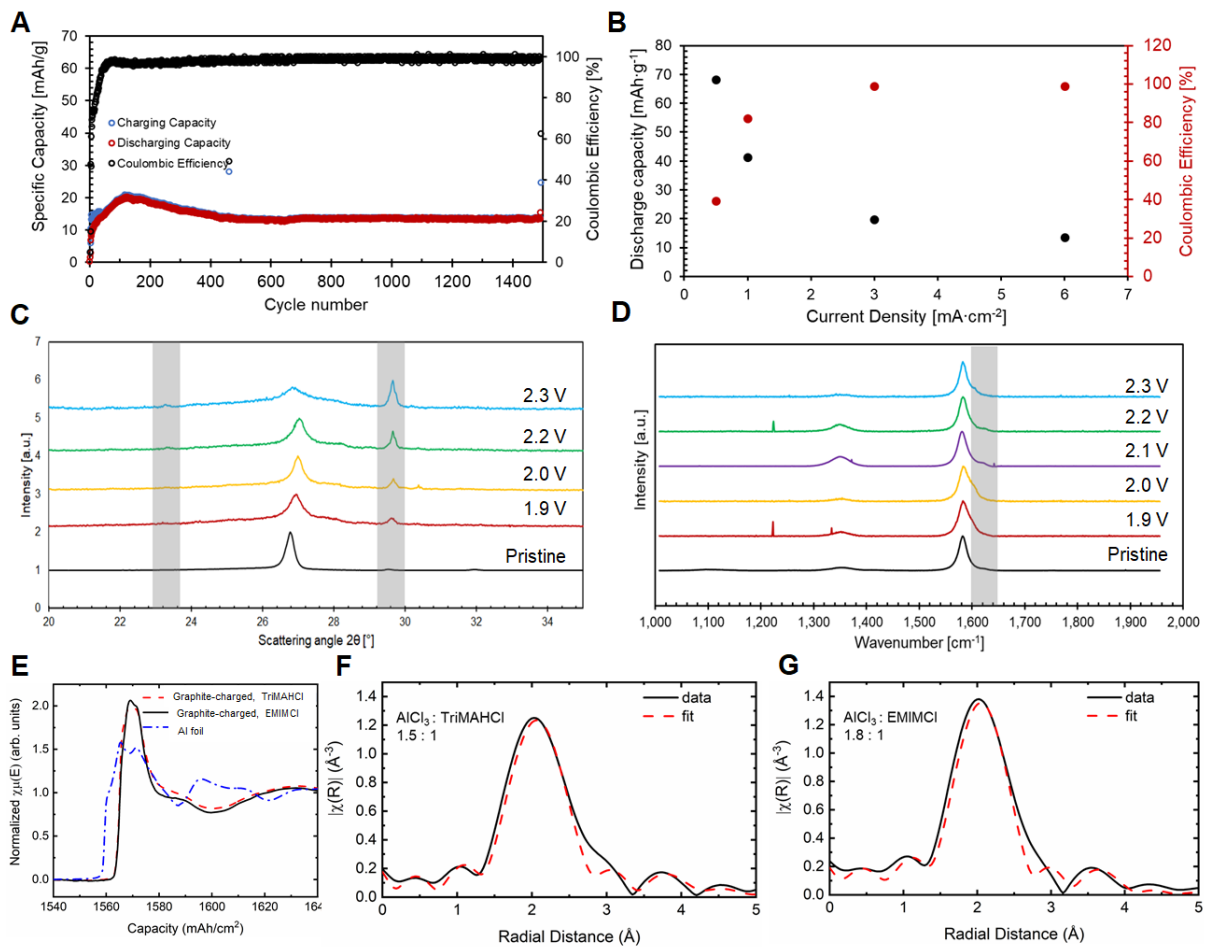

**Figure 6.** Electrochemical cycling behavior of Al-graphite cells and materials characterization of graphite cathode (**A**) Long-term stability cycling test of an Al-graphite cell over 1300 charging and discharging cycles at a current density of 335.4 mA·g$^{-1}$ (6 mA·cm$^{-2}$) using 1.5: 1 AlCl$_3$-TriMAHCl (trimethylamine hydrochloride) as the electrolyte (**B**) Specific discharge capacity and Coulombic efficiency as a function of current density of Al-graphite cells (**C**) X-ray diffraction on graphite cathodes upon charging at 1 mA·cm$^{-2}$ using 1.5: 1 AlCl$_3$-TriMAHCl (trimethylamine hydrochloride) as the electrolyte (**D**) Raman on graphite cathodes upon charging at 1 mA·cm$^{-2}$ using 1.5: 1 AlCl$_3$-TriMAHCl (trimethylamine hydrochloride) as the electrolyte (**E**) XANES spectra of charged graphite cathodes and an Al foil standard (**F**) k$^2$-weighted |χ(R)| and corresponding EXAFS fit for graphite electrode charged in AlCl$_3$: TriMAHCl (1.5:1) (**G**) k$^2$-weighted |χ(R)| and corresponding EXAFS fit for graphite electrode charged in AlCl$_3$: EMIMCl (1.8:1). Fitting for F and G was

performed using a theoretical model for the tetrachloroaluminate ion, $AlCl_4^-$. R-factors for the fits were 3.1 % and 3.7 % for (F) and (G) respectively. (See also Figures S17-20, Table S1).

Fig. 6A reports the discharge capacity and columbic efficiency of Al *vs.* graphite full cells cycled at a fixed current density of 335.4 mA·g$^{-1}$ (6 mA·cm$^{-2}$). Under these conditions, a specific discharge capacity of approximately 15 mAh·g$^{-1}$ (based on the graphitic carbon mass in the cathode) and average Coulombic efficiency of 97.65% are maintained for over 1,300 cycles in battery cells that use $AlCl_3$-TriMAHCl (1.5: 1) as electrolyte. While lower than the approximately 60 mAh·g$^{-1}$ reported by Lin *et al.* (*2*) using a custom fabricated type of graphitic carbon foam as the cathode, our findings are in good agreement with expectations for a stage-3 graphite intercalation compound (GIC) based on analysis of the diffraction patterns (Fig 6C) and theoretical predictions (*34*) . Bhauriyal *et al.*(*34*) predicted *via* first-principles calculations that graphite can store from 25.94 to 69.62 mAh g$^{-1}$ of chloroaluminate species for a stage-4 and stage-1 GIC, respectively; implying that with better cathode design and lower discharge rates, higher discharge capacities are achievable, as seen in Fig 6B. The current cathode design can achieve a stage-1 GIC, reaching the theoretical value predicted by Bhauriyal *et al.*, however, the reversibility of the process is reduced significantly. Additionally, the intercalant gallery height (distance separating two graphite layers) was estimated to be ~5.22 Å based on the diffraction patterns, suggesting that the $AlCl_4^-$ anions with a comparable size (~5.28 Å) (*35*) are the most likely intercalant species.

Prior experimental (*2*) and theoretical work (*36, 37*) have proposed that a distorted tetrahedral geometry of $AlCl_4^-$ is intercalated as opposed to a planar geometry (*38*) given its higher thermodynamic stability. The distortion results from the van der Waals forces between the graphite layers, reducing the graphite interlayer distance by compressing the size of tetrahedral $AlCl_4$, giving it a distorted geometry. (*34*) The Raman spectra collected on the same cathodes was also performed to probe chloroaluminate anion intercalation into graphite upon charge (Fig. 6D). The graphite G band (~1584 cm$^{-1}$) shows evidence of a right shoulder at ~1605 cm$^{-1}$ upon anion intercalation, in good agreement with prior work by Lin *et al.* and Angell *et al.* (*2, 6*). We note that the initial G band remains largely intact, indicating that more hosting capacity exists in the cathode than utilized in the battery discharge when using 1 mA·cm$^{-2}$, as it was observed when charging at a lower current density (0.5 mA·cm$^{-2}$), and achieving a larger gravimetric capacity (Fig. 6B).

The XRD and Raman measurements both suggest that anion intercalation occurs upon charging and that $AlCl_4^-$ is the intercalation species in the graphite cathode. Nevertheless, to more conclusively establish the identity of the intercalated species, we performed X-ray Absorption Spectroscopy (XAS) and Al-edge extended X-ray absorption fine structure (EXAFS) measurements. XAS measurements were collected on graphite cathodes charged to 2.3 V where the electrolyte was either $AlCl_3$: TriMAHCl (1.5:1) or $AlCl_3$: EMIMCl (1.8:1), for comparison. The X-ray Absorption Near-Edge Structure (XANES) region of the XAS spectra of the charged samples as well as an Al metal foil standard are shown in Fig 6E. The edge energy of the charged samples was approximately 5.2 eV higher than the foil standard, indicating an oxidized Al species at the electrode. The sharp edge arises from Al $1s \rightarrow 3p$ transitions and is ~ 2 eV lower in energy than reported transitions edge energy for corundum $Al_2O_3$. (*39*)

EXAFS analysis is well suited for understanding the structure of local atomic low crystallinity or non-crystalline materials lacking long-range order, as it is sensitive to the local atomic structure within the first several coordination shells around the central absorbing atom (*40, 41*). The Fourier transformed EXAFS spectra of the charged graphite electrodes with the $AlCl_3$-TriMAHCl and $AlCl_3$-EMIMCl electrolytes are shown in Fig 6F, G. The spectra were successfully fit using a theoretical model derived from the reported crystal structure of a lithium tetrachloroaluminate salt, (37) with an Al atom tetrahedrally surrounded by four chlorine atoms (fig S19A). The model utilizes three distinct photoelectron scattering paths – two single scattering paths corresponding to Al-Cl bonds with nominal interatomic distance of 2.10 Å and 2.14 Å, and a multiple scattering path at 3.90 Å. The coordination numbers were set to the crystallographic values. The experimental data fit well to the model, with R-factors of 3.1 and 3.7 for samples charged in $AlCl_3$: TriMAHCl (1.5:1) and AlCl$_3$: EMIMCl (1.8:1) electrolytes, respectively. Modeled interatomic distances of 2.09 – 2.10 Å for the single scattering paths and 3.90 Å for the multiple scattering path are in excellent agreement with the reported crystal structure. Debye Waller disorder factor ($\sigma^2$) values were determined to range from 0.011 to 0.012 Å$^2$. Full EXAFS fitting results are presented in table S1.

An alternative fitting model was also attempted that was derived from the crystal structure of a heptachloroaluminate salt. (*42*) Heptachloroaluminate ions are comprised of two AlCl$_4$ tetrahedra connected via a chlorine atom in a bent corner sharing configuration, with three Al-Cl bonds at 2.09 – 2.13 Å and a fourth, longer path at 2.22 - 2.26 Å (fig S19B). (*42*) Additional scattering paths are also predicted for non-bonding Al-Cl and Al-Al interactions at 3.74 Å and 3.69 Å, respectively. Fitting of the experimental data to the heptachloroaluminate model resulted in a shift in interatomic radial distances of the first shell Al-Cl paths to lower values than predicted by the heptachloroaluminate crystal structure such that the average Al-Cl distance was 2.11 Å, which is the nominal first coordination shell distance for the tetrachloroaluminate structure. The Al-Cl and Al-Al paths also shifted from their nominal values by ~0.10 Å to 3.86 Å, which is close to the radial distance of the multiple scattering path in AlCl$_4^-$. Furthermore, if the interatomic distances were set to the nominal values from the published crystal structure, greater misfit was observed (fig S20) when compared to the fits based on the tetrachloroaluminate structure (R-factor of 6.3 – 7.0% for heptachloroaluminate fitting model compared with 3.1 – 3.7% for tetrachloroaluminate fitting model). These results provide evidence that the electroactive species at the cathode is the tetrachloroaluminate ion rather than the heptachloroaluminate ion. These results therefore motivate the design of cathode materials that allow fast diffusion of bulky ions, show reversible phase transformations and maintain structural integrity upon cycling.

Conclusions

We report that low-cost electrolytes based on quaternary ammonium-based salts of broken symmetry can be designed with melting points below room temperature. By means of electrochemical, spectroscopic, and morphological analyses we evaluate the critical role of each chloroaluminate species in Al electrochemical properties in batteries. We find that the reversibility of Al plating and stripping, the nature of the solid electrolyte interphase it forms, and the morphological evolution during plating and stripping of Al depend sensitively and quantitatively on the ratio of AlCl$_4^-$ to Al$_2$Cl$_7^-$ ions in the electrolyte. The sensitivity stems from the dual role the ions play in facilitating electroreduction at the Al/electrolyte interface and in etching the resistive alumina surface layer that forms on the Al that prevents transport to the interface. A key finding is that provided the ratio of AlCl$_4^-$ to Al$_2$Cl$_7^-$ ions can be preserved in room temperature electrolytes with broken cation symmetry, it is possible to achieve highly efficient plating and stripping of Al in cost-effective quaternary ammonium-based electrolyte media.

We leverage the last discovery to create Al||graphite electrochemical cells and study them as platforms for achieving low-cost, long-duration storage of electrical energy. Galvanostatic charge-discharge measurements show that these cells demonstrate stable long-term cycling performance, particularly when AlCl$_3$-TriMAHCl (1.5:1) is used as electrolyte. The electrolyte molecular design approach presented here therefore offers a promising, new route towards achieving low melting point electrolytes with the desired Lewis acidity for highly reversible Al electrodeposition and dissolution in secondary batteries.

Supplemental information description

Supplemental figures 1 – 20

Supplemental table 1

Supplemental experimental procedures

Supplemental references

Acknowledgements

The authors would like to thank Ivan Keresztes for valuable discussions regarding NMR data acquisition and analysis.

The cross-section imaging of Al electrodes was conducted in the MIT.nano Characterization Facilities by JZ using Raith VELION FIB-SEM system and Themis STEM.

The XAS and EXAFS measurements were conducted at Brookhaven National Laboratory by DCB, CJ, DAF.

Funding:
- Center for Mesoscale Transport Properties, an Energy Frontier Research Center supported by the U.S. Department of Energy, Office of Science, Basic Energy Sciences, under award no. DESC0012673.
- RGM acknowledges that this study is based on work supported by a Cornell Energy Systems Institute Post-Doctoral Fellowship.
- Cornell Center for Materials Research Shared Facilities are supported through the NSF MRSEC program (DMR-1719875).
- Cornell University NMR Facility, which is supported, in part, by the NSF through MRI award CHE-1531632.

Author contributions

Conceptualization: RGM, JZ, LAA. Formal Analysis: RGM. Data Curation and Methodology: RGM. Investigation: RGM, JZ, DCB, CJ, DAF. Visualization: RGM. Funding acquisition: RGM, ACM, EST, LAA. Project administration: RGM, JZ, LAA. Supervision: ACM, EST, LAA. Validation: RGM. Writing – original draft: RGM. Writing – review & editing: RGM, JZ, DCB, CJ, DAF, ACM, JKT, EST, LAA.

Declaration of interests

RGM, JZ, and LAA have filed a provisional US patent related to this work, application number 63/321,747.

# Supplemental experimental procedures

Supplemental items

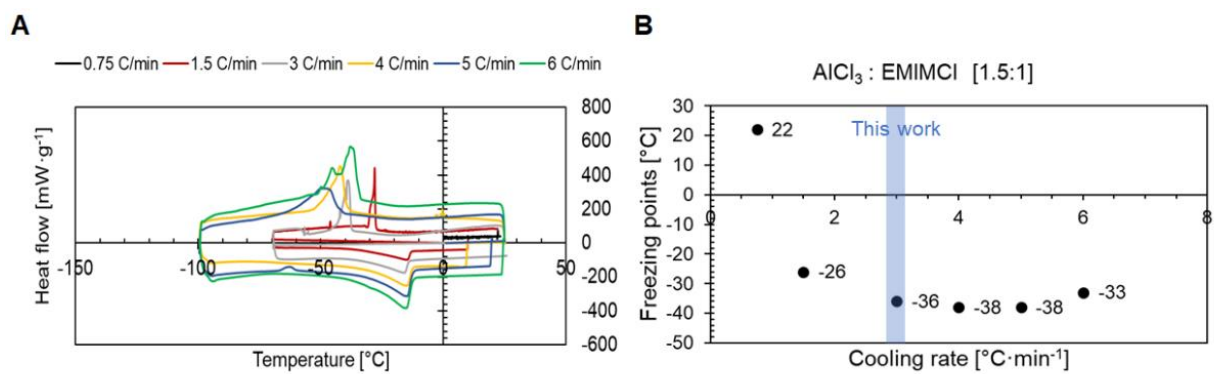

**Figure S1.** Effect of cooling rate on freezing points in AlCl$_3$-[EMIM]Cl (1.5:1, in molar ratio). (**A**) Differential Scanning Calorimetry (DSC) spectra with varying cooling rates from 0.75 to 6 °C·min$^{-1}$ (**B**) Freezing points as a function of cooling rate, related to Figure 2.

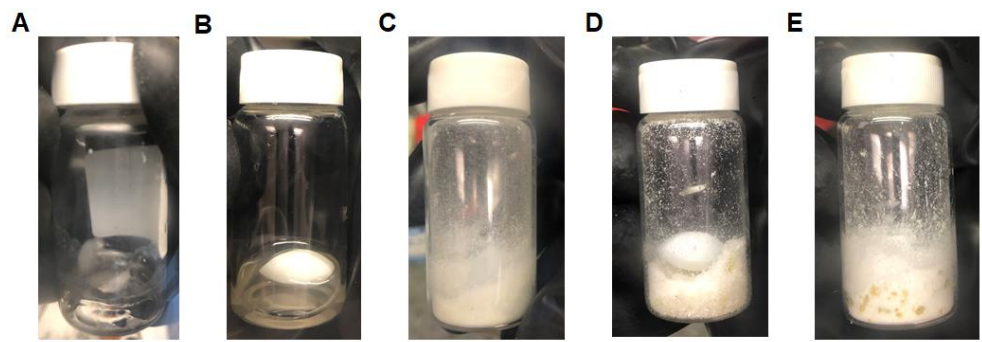

**Figure S2.** Images of AlCl$_3$ -ammonium-based compounds (molar ratio 1.5: 1) (**A**) Trimethylamine-HCl (**B**) Triethylamine -HCl (**C**) Tetramethylammonium chloride (**D**) Tetraethylammonium chloride (**E**) Tetrapropylammonium chloride. Images B and D show the magnetic stirrer inside the glass vial, related to Figure 2.

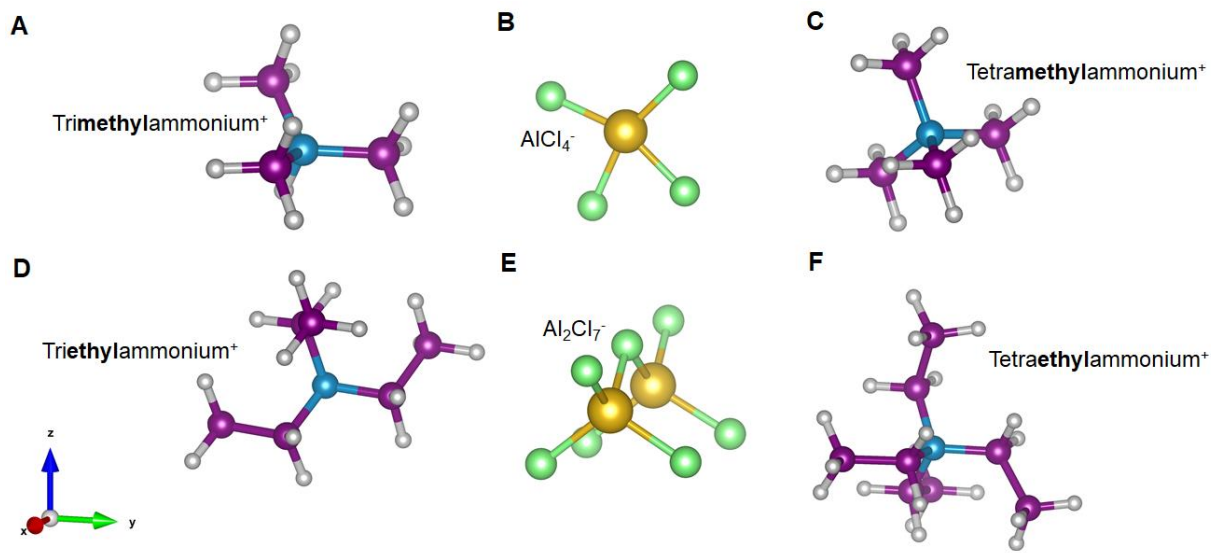

**Figure S3.** Molecular structures of cations and anions present in the ammonium-based electrolytes. Blue = Nitrogen, Purple = Carbon, Gray = Hydrogen, Yellow = Aluminum, Green = Chlorine. (**A**) Trimethylammonium$^+$ (TriMAH$^+$), (**B**) Tetrachloroaluminate$^-$ (AlCl$_4^-$), (**C**) Tetramethylammonium$^+$ (TetraMA$^+$), (**D**) Triethylammonium$^+$ (TriEAH$^+$), (**E**) Heptachloroaluminate (Al$_2$Cl$_7^-$), (**F**) Tetraehtylammonium$^+$ (TetraEA$^+$).

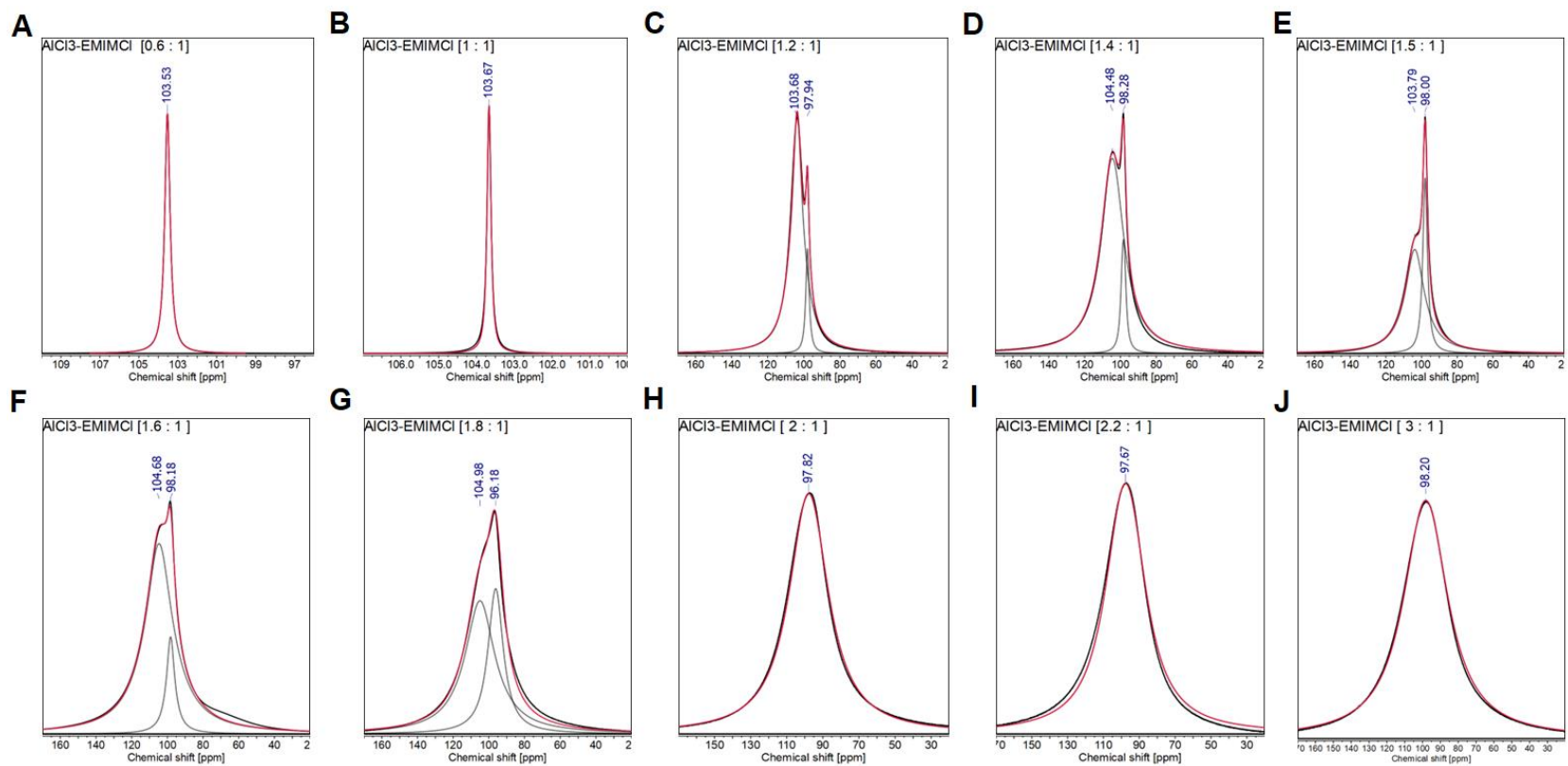

**Figure S4.** Quantitative deconvolution of AlCl$_4^-$ and Al$_2$Cl$_7^-$ via liquid-state $^{27}$Al NMR on AlCl$_3$-[EMIM]Cl ionic liquid electrolyte mixtures, varying AlCl$_3$ molar concentration keeping 1 mole of [EMIM]Cl constant (**A**) 0.6 AlCl$_3$, (**B**) 1 AlCl$_3$, (**C**) 1.2 AlCl$_3$, (**D**) 1.4 AlCl$_3$, (**E**) 1.5 AlCl$_3$, (**F**) 1.6 AlCl$_3$, (**G**) 1.8 AlCl$_3$, (**H**) 2 AlCl$_3$, (**I**) 2.2 AlCl$_3$ and (**J**) 3 AlCl$_3$. Traces; black – raw data, blue – fitting for individual species AlCl$_4^-$ and Al$_2$Cl$_7^-$, red – fitting of the sum of species, orange – residual between raw data and fitting, related to Figure 3A.

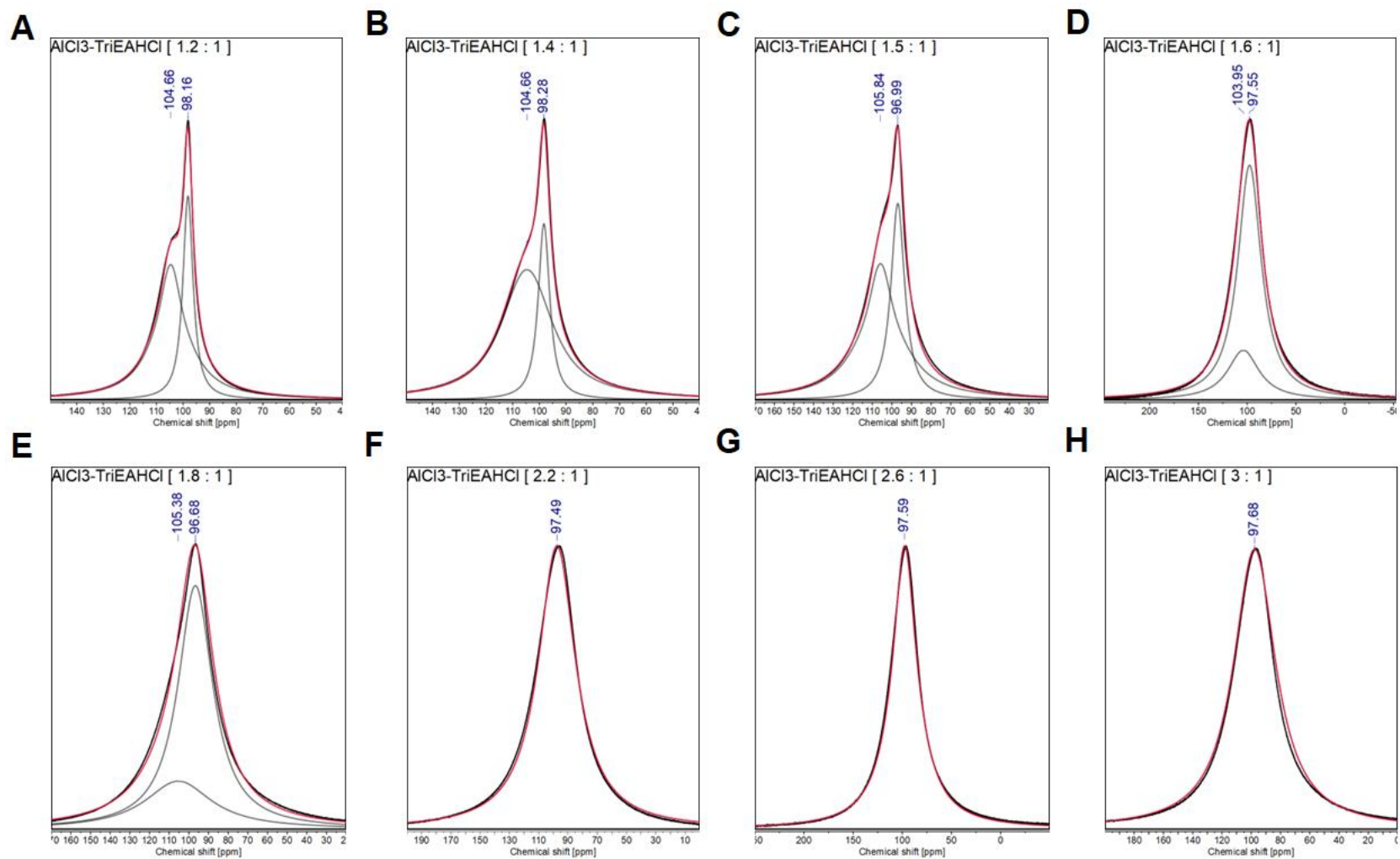

**Figure S5.** Quantitative deconvolution of AlCl$_4^-$ and Al$_2$Cl$_7^-$ via liquid-state $^{27}$Al NMR on AlCl$_3$-TriEAHCl ionic liquid electrolyte mixtures, varying AlCl$_3$ molar concentration keeping 1 mole of TriEAHCl constant (**A**) 1.2 AlCl$_3$, (**B**) 1.4 AlCl$_3$, (**C**) 1.5 AlCl$_3$, (**D**) 1.6 AlCl$_3$, (**E**) 1.8 AlCl$_3$, (**F**) 2.2 AlCl$_3$, (**G**) 2.6 AlCl$_3$, (**H**) 3 AlCl$_3$. Traces; black – raw data, blue – fitting for individual species AlCl$_4^-$ and Al$_2$Cl$_7^-$, red – fitting of the sum of species, orange – residual between raw data and fitting, related to Figure 3B.

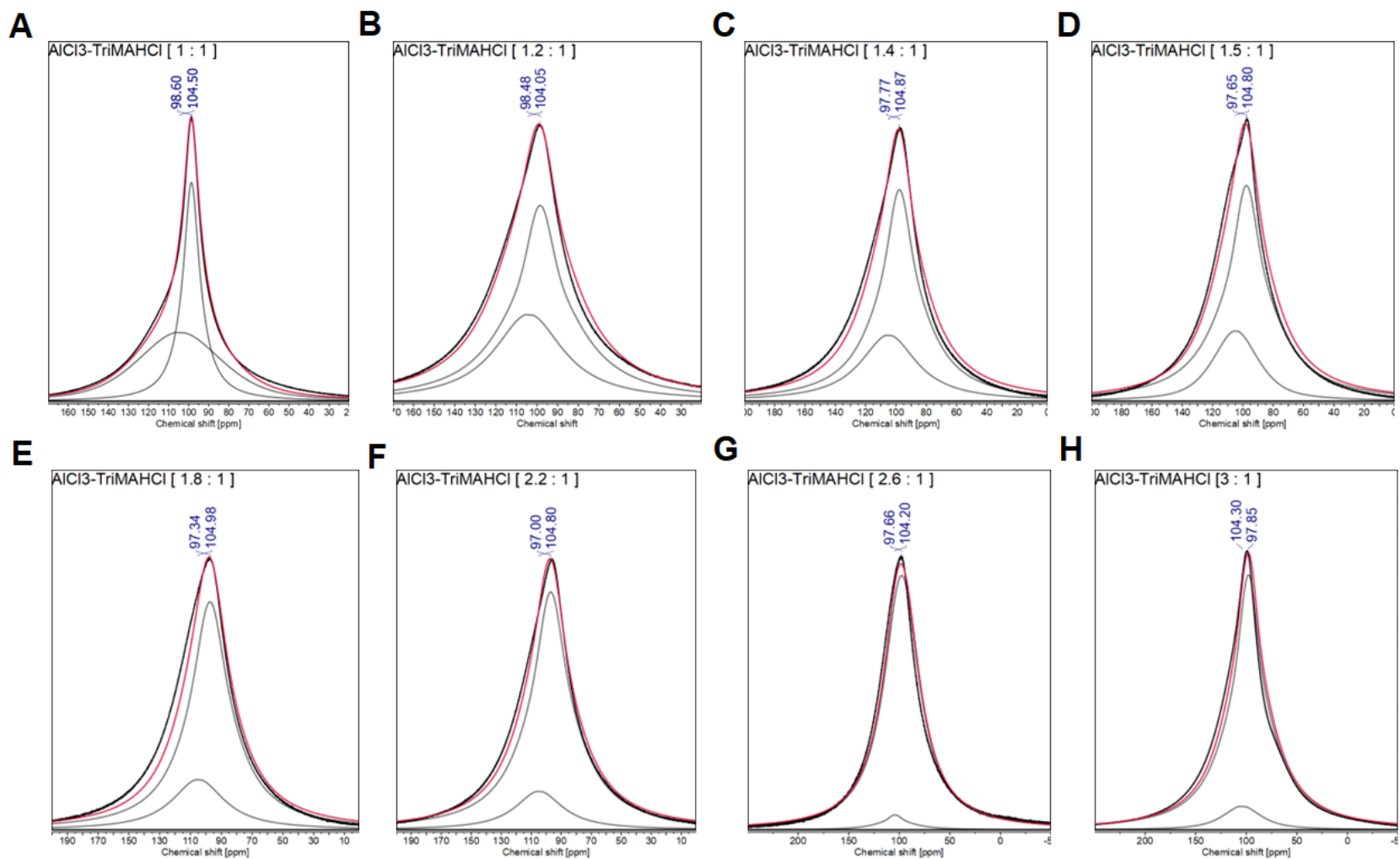

**Figure S6.** Quantitative deconvolution of $AlCl_4^-$ and $Al_2Cl_7^-$ via liquid-state $^{27}Al$ NMR on $AlCl_3$-TriMAHCl ionic liquid electrolyte mixtures, varying $AlCl_3$ molar concentration keeping 1 mole of TriMAHCl constant (**A**) 1 $AlCl_3$, (**B**) 1.2 $AlCl_3$, (**C**) 1.4 $AlCl_3$, (**D**) 1.5 $AlCl_3$, (**E**) 1.8 $AlCl_3$, (**F**) 2.2 $AlCl_3$, (**G**) 2.6 $AlCl_3$, (**H**) 3 $AlCl_3$. Traces; black – raw data, blue – fitting for individual species $AlCl_4^-$ and $Al_2Cl_7^-$, red – fitting of the sum of species, orange – residual between raw data and fitting, related to Figure 3C.

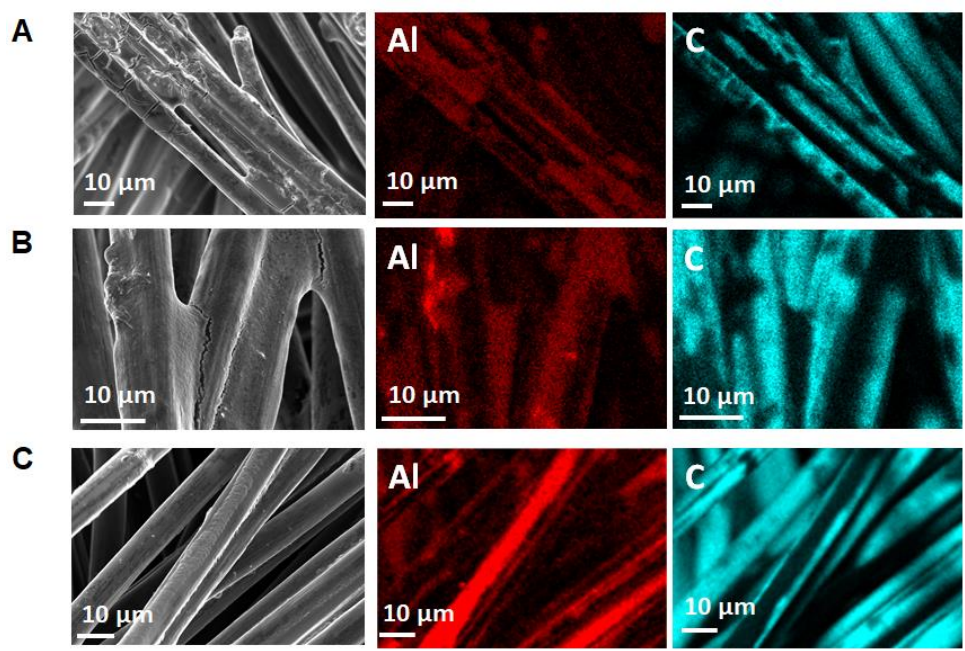

**Figure S7.** Aluminum electrodeposition in room temperature molten salts (1.5: 1, in molar ratio - $AlCl_3$: salt) onto carbon cloth substrate electrodes, SEM, and corresponding EDS elemental maps (**A**) $AlCl_3$-[EMIM]Cl, (**B**) $AlCl_3$-[TriMAH]Cl, and (**C**) $AlCl_3$-[TriEAH]Cl, related to Figure 4.

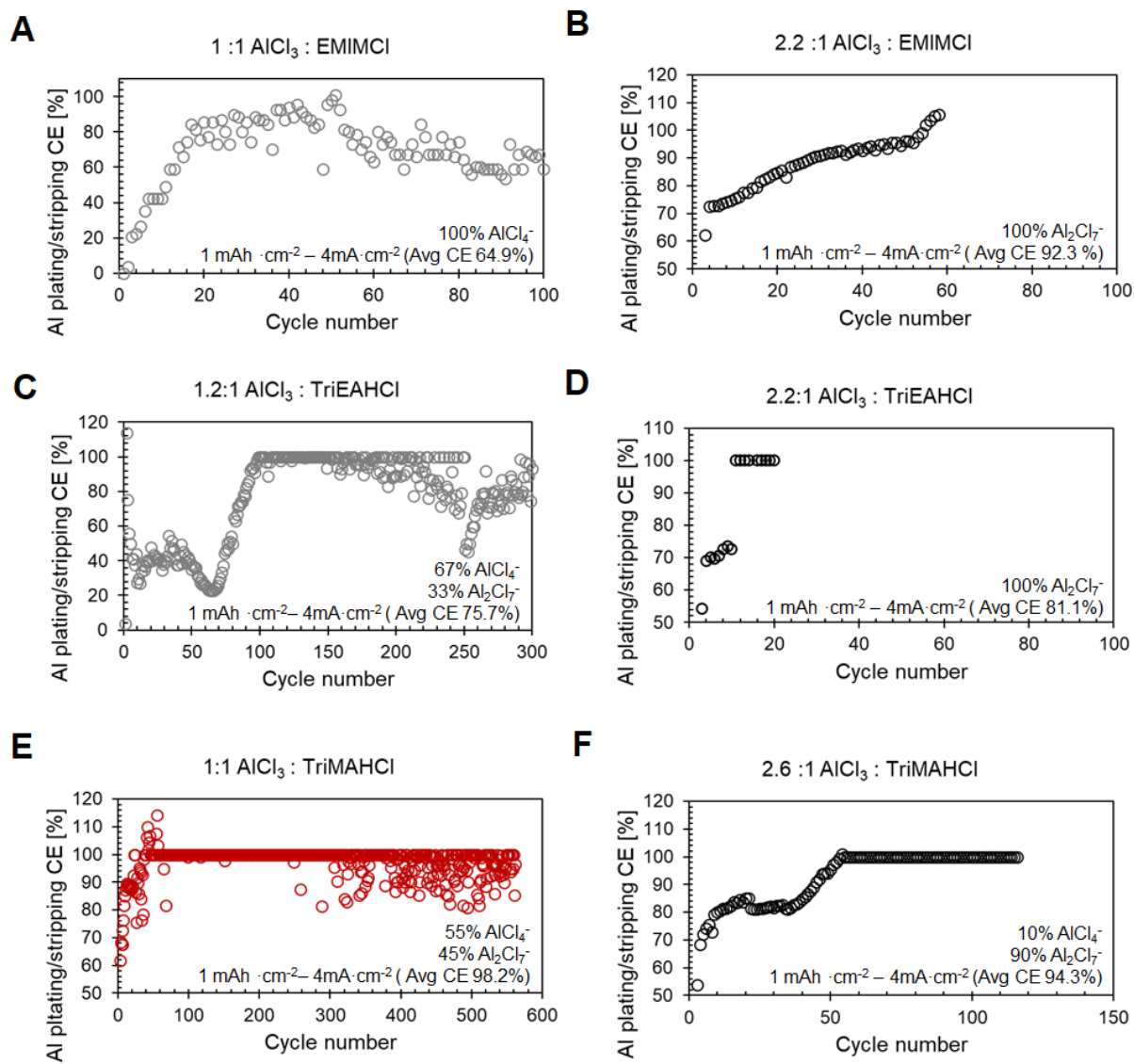

**Figure S8.** Electrochemical cycling behavior of Al electrodes in galvanostatic plating/stripping experiments using various electrolytes. (**A, B**) 1: 1 and 2.2: 1 AlCl$_3$-[EMIM]Cl (1-ethyl-3-methyl-imidazolium chloride), respectively, (**C, D**) 1.2: 1 and 2.2: 1 AlCl$_3$-TriEAHCl (Triethylamine hydrochloride), respectively, (**E, F**) 1: 1 and 2.6: 1 AlCl$_3$-TriMAHCl (Trimethylamine hydrochloride), respectively, related to Figure 4.

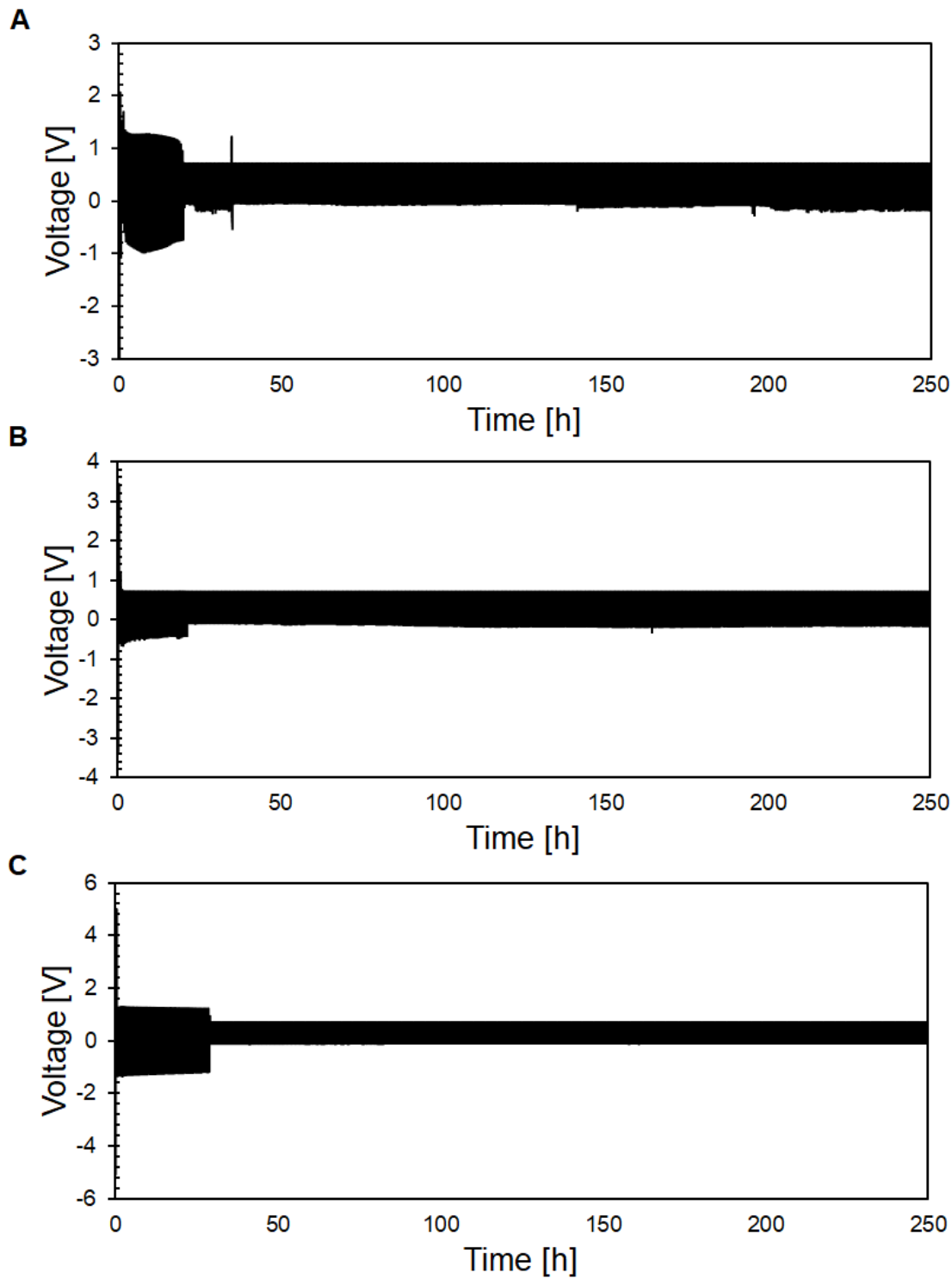

**Figure S9.** Voltage response in Al symmetric cells under galvanostatic cycling at 1mAhcm-2 areal capacity, 4 mAcm-2 using different electrolytes. (**A**) 1.8 : 1 AlCl$_3$ – EMIMCl (1-ethyl-3-methyl-imidazolium chloride), (**B**) 1.6:1 AlCl$_3$-TriEAHCl (Triethylamine hydrochloride), (**C**) 1.5 : 1 AlCl$_3$-TriMAHCl (Trimethylamine hydrochloride), related to Figure 4.

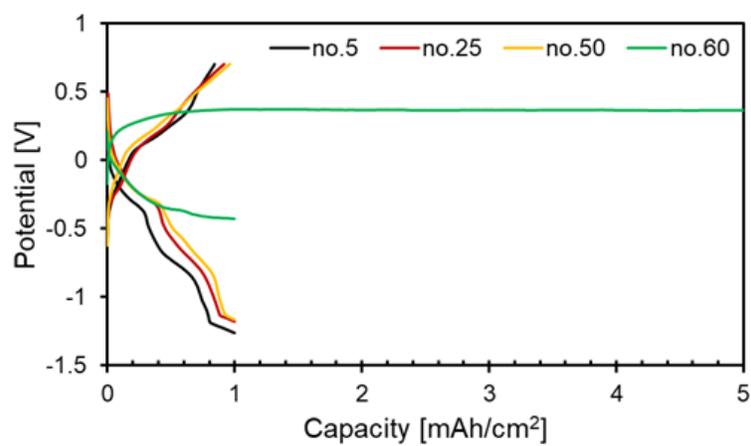

**Figure S10.** Representative potential curves obtained during Al plating/stripping at 1mAh·cm$^{-2}$ areal capacity, 4 mA·cm$^{-2}$ demonstrate electrochemical corrosion occurring in the cells when using 2.6:1 AlCl$_3$-TriMAHCl (Trimethylamine hydrochloride) electrolyte, related to Figure 4.

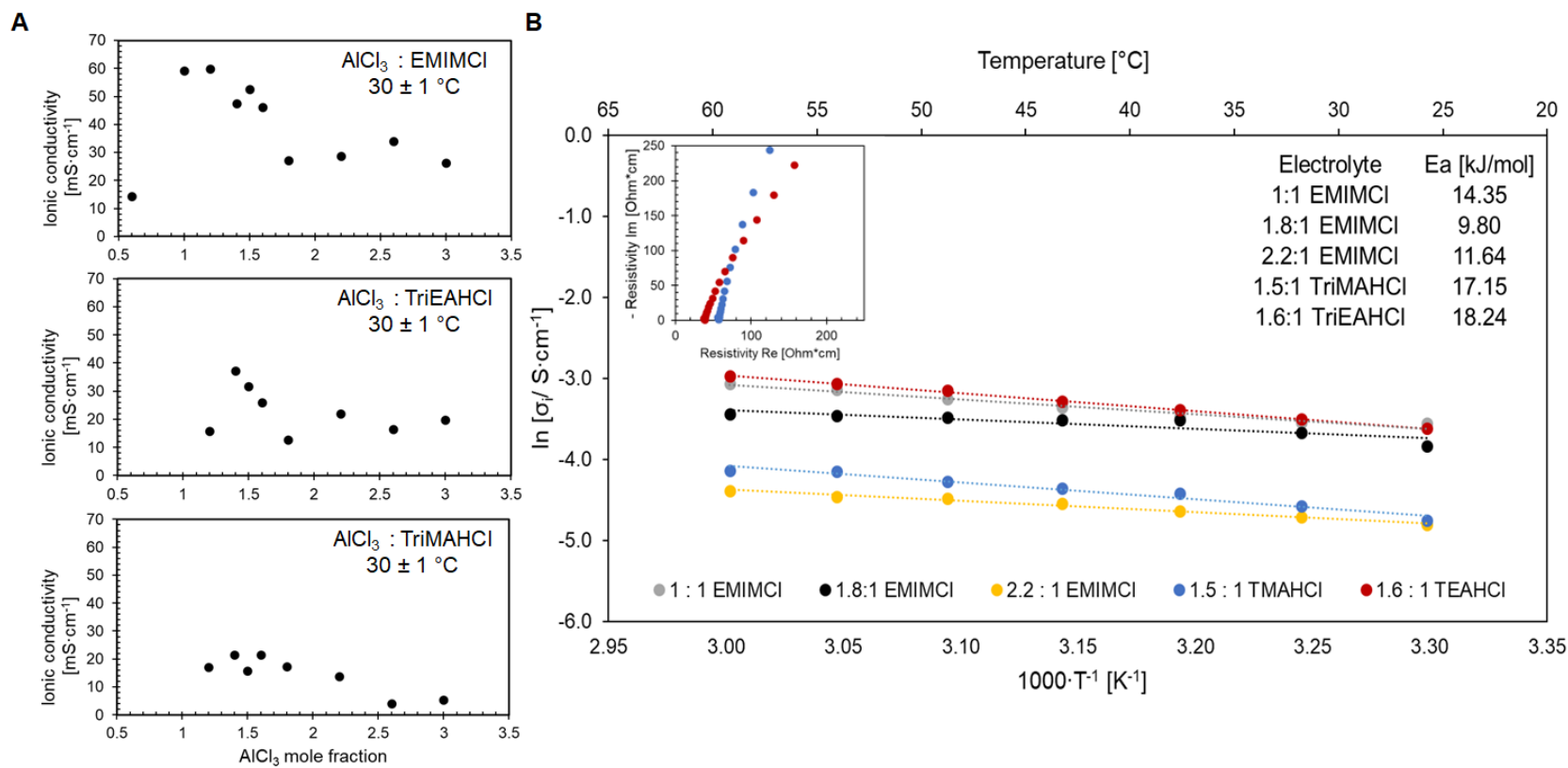

**Figure S11.** Ionic transport of imidazolium- and ammonium-based electrolytes as a function of AlCl$_3$ mole fraction and temperature. (**A**) Ionic conductivity values measured as a function of AlCl$_3$ mole fraction at 30°C via electrochemical impedance spectroscopy. (top) AlCl$_3$-[EMIM]Cl, (middle) AlCl$_3$-TriEAHCl, (bottom) AlCl$_3$-TriMAHCl (**B**) Ionic conductivity as a function of temperature and calculated energy of activation for transport. Inset shows two representative Nyquist plots used to determine the ionic conductivity of the liquid electrolytes.

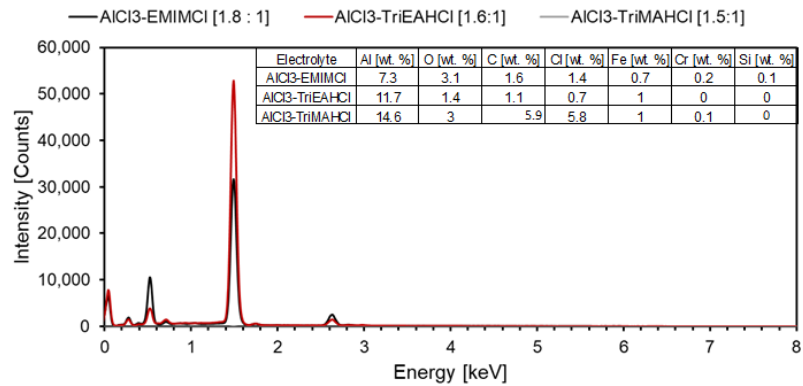

**Figure S12.** Energy Dispersive Spectroscopy (EDS) spectra of Al anodes after 100 cycles of plating and stripping, ending with a plating step, immersed in different electrolytes, related to Figure 5.

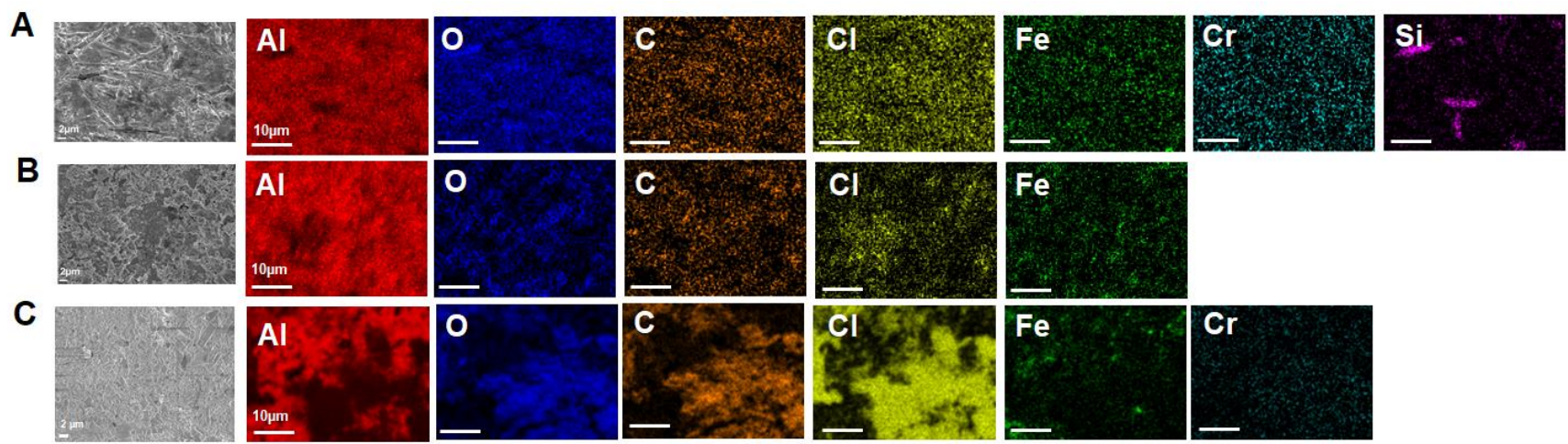

**Figure S13.** Energy Dispersive Spectroscopy (EDS) elemental mapping of Al anodes after 100 cycles of stripping and plating, ending with a plating step using different electrolytes. (**A**) 1.8:1 $AlCl_3$-EMIMCl (**B**) 1.6:1 $AlCl_3$-TriEAHCl (**C**) 1.5:1 $AlCl_3$-TriMAHCl, related to Figure 5.

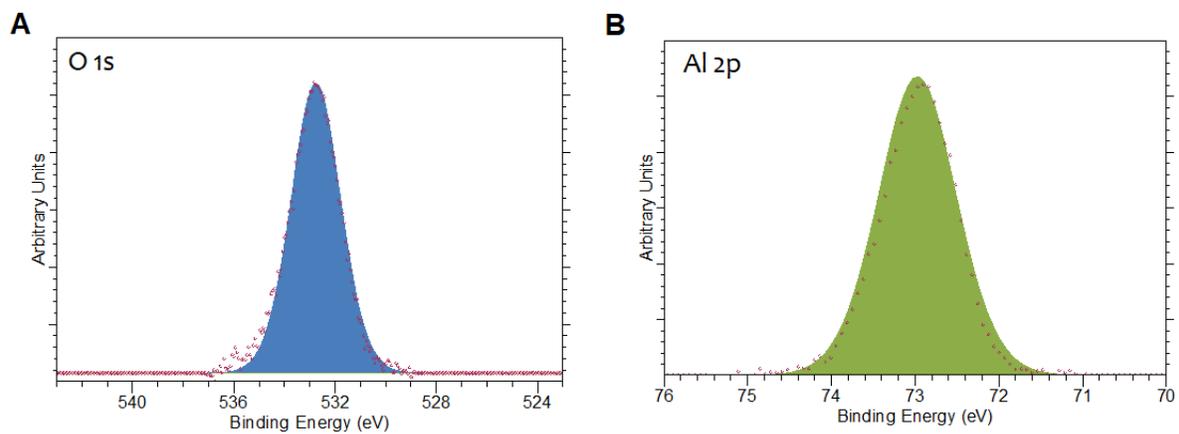

**Figure S14.** X-Ray Photoelectron Spectroscopy (XPS) analysis of Al foil as received. (**A**) O 1s core level (**B**) Al 2p core level, related to Figure 5.

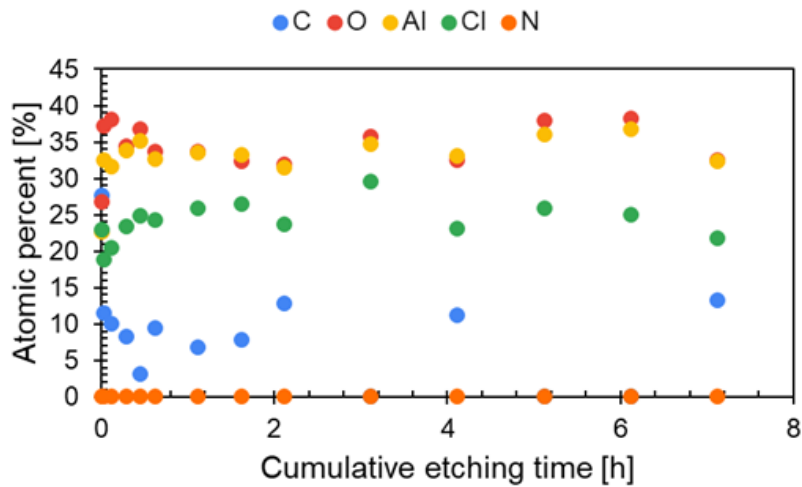

**Figure S15.** Variation in atomic percent of C, O, Al, Cl, N as a function of Ar-sputtering time (etching time) of Al anodes collected after galvanostatic plating/stripping in 1.5: 1 $AlCl_3$-[TriMAH]Cl (Trimethylamine hydrochloride) electrolyte, determined *via* x-ray photoelectron spectroscopy (XPS) survey scans, related to Figure 5.

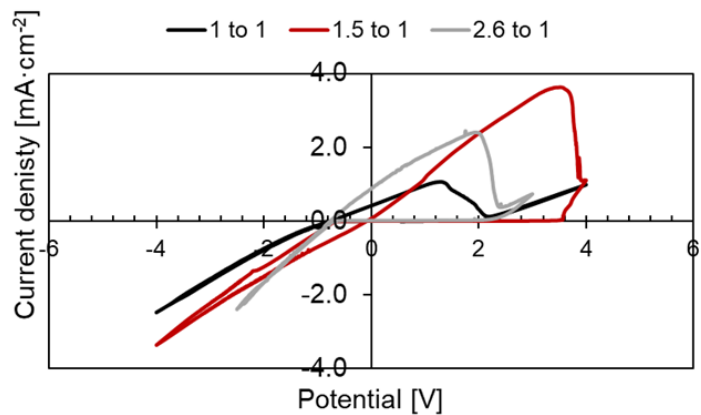

**Figure S16.** Three-electrode cyclic voltammetry of $AlCl_3$-TriMAHCl electrolytes as a function of mole concentration of $AlCl_3$. Glassy carbon was used as the working electrode, Al as the counter electrode and Pt as the reference electrode.

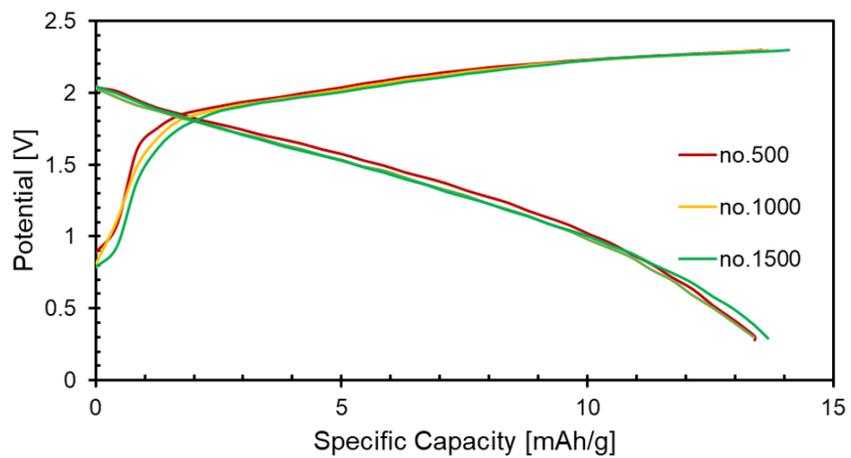

**Figure S17.** Galvanostatic charge and discharge curves of an Al-graphite cell at a current density of 335.4 mA·g-1 (6 mA·cm-2) using 1.5: 1 $AlCl_3$-TriMAHCl (trimethylamine hydrochloride) as the electrolyte, related to Figure 6.

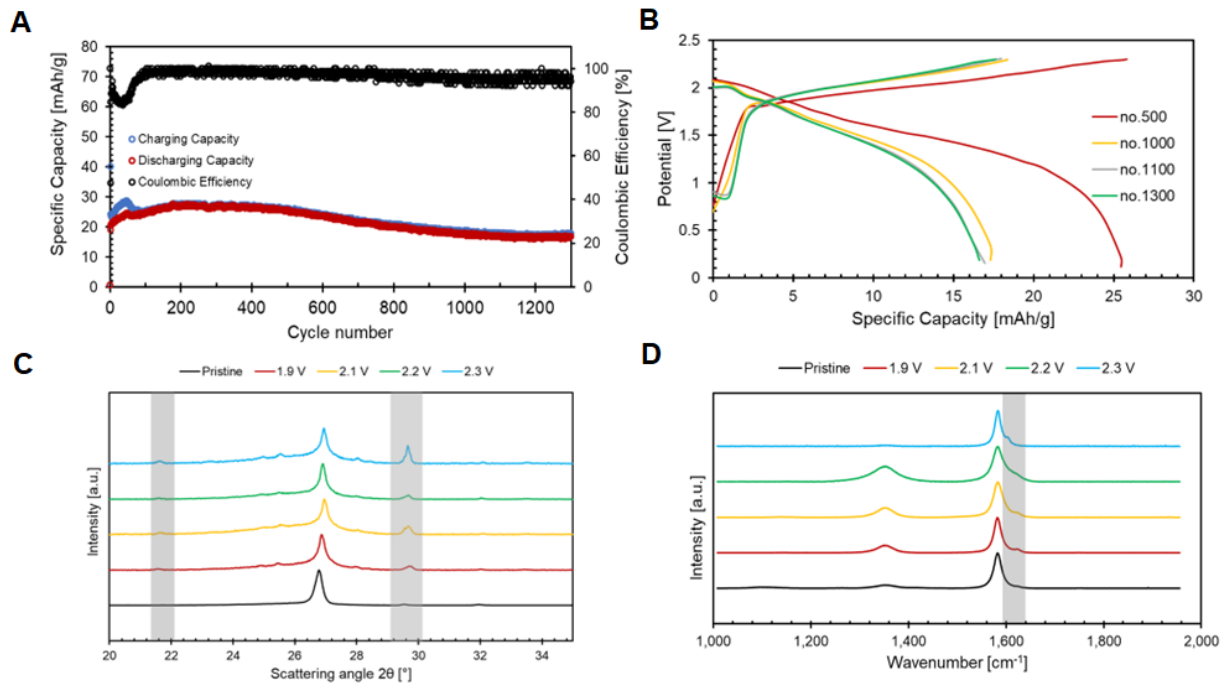

**Figure S18.** Electrochemical cycling behavior of Al-graphite cells and materials characterization of graphite cathode using 1.6: 1 AlCl$_3$-TriEAHCl (Triethylamine hydrochloride) electrolyte. (**A**) Long-term cycling stability test of an Al-graphite cell for 1300 charging and discharging cycles at a current density of 335.4 mA·g$^{-1}$ (6 mA·cm$^{-2}$) (**B**) Galvanostatic charge and discharge curves from A (**C**) X-ray diffraction on graphite cathodes upon charging at 1 mA·cm$^{-2}$ (**D**) Raman on graphite cathodes upon charging at 1 mA·cm$^{-2}$.

**Table S1.** EXAFS fitting results of graphite cathodes. Uncertainties in the interatomic distances, energy shift parameter ($E_0$) values and Debye-Waller factor ($\sigma^2$) were determined from EXAFS fitting. A detailed description of how the uncertainties here are calculated can be found in Ref. (*5*), related to Figure 6.

| Sample | R-factor (%) | Path | $S_0^2$ | Coordination Number | Nominal Path Length (Å) | Interatomic distance (Å) | $E_0$ | $\sigma^2$ (x $10^{-2}$ Å$^2$) |
|---|---|---|---|---|---|---|---|---|
| Graphite, 2.3 V, AlCl3: EMIMCl 1.8:1 | 3.7 | Al-Cl (single scattering) | 0.88 | 3 | 2.10 | 2.08 ± 0.04 | -2 ± 3 | 1.2 ± 0.2 |
| | | Al-Cl (single scattering) | 0.88 | 1 | 2.14 | 2.11 ± 0.04 | -2 ± 3 | 1.2 ± 0.2 |
| | | Al-Cl-Cl (obtuse triangle) | 0.88 | 8 | 3.90 | 3.90 ± 0.01 | -2 ± 3 | 1.1 ± 0.2 |
| Graphite, 2.3 V, AlCl3: TMAHCl 1.5:1 | 3.1 | Al-Cl (single scattering) | 0.88 | 3 | 2.10 | 2.09 ± 0.03 | -2 ± 2 | 1.1 ± 0.2 |
| | | Al-Cl (single scattering) | 0.88 | 1 | 2.14 | 2.09 ± 0.03 | -2 ± 2 | 1.1 ± 0.2 |
| | | Al-Cl-Cl (obtuse triangle) | 0.88 | 8 | 3.90 | 3.89 ± 0.04 | -2 ± 2 | 1.1 ± 0.2 |

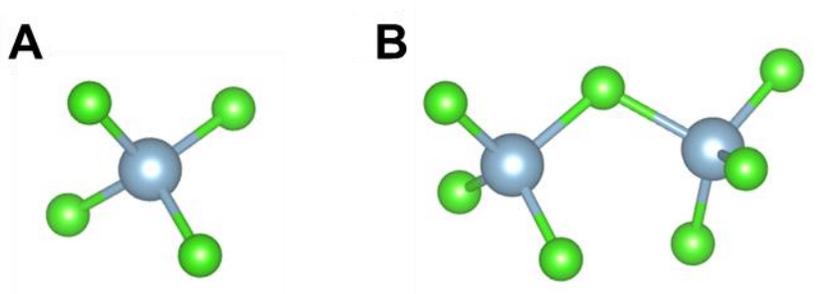

**Figure S19.** Theoretical structures of (**A**) AlCl$_4^-$ and (**B**) Al$_2$Cl$_7^-$ used for EXAFS fitting, related to Figure 6.

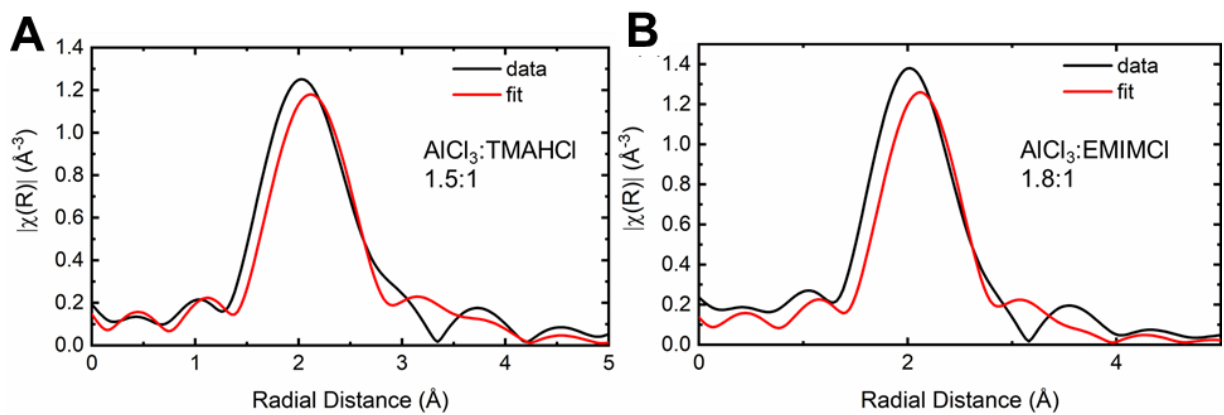

**Figure S20.** k2-weighted |χ(R)| and corresponding EXAFS fit graphite electrodes charged in (**A**) AlCl$_3$:TriMAHCl (1.5:1) or (**B**) AlCl$_3$:EMIMCl. Fitting was performed using a theoretical model for heptachloroaluminate, Al$_2$Cl$_7^-$. The R-factor for the fits were 6.3% and 7.0% for (**A**) and (**B**), respectively, related to Figure 6.

Supplemental experimental procedures

**Electrolyte preparation**
The chloroaluminate compounds were prepared by slowly adding appropriate amounts of anhydrous AlCl$_3$ to 1-ethyl-3-methylimidazolium chloride (EMIMCl), trimethylamine-hydrochloride (TriMAHCl), triethylamine-hydrochloride (TriEAHCl), tetramethylammonium chloride (TetraMACl) or tetraethylammonium chloride (TetraEACl) while stirring in a dry Ar-filled glovebox (<0.1 ppm H$_2$O, < 1 ppm O$_2$).

**Phase diagram construction**
Differential Scanning Calorimetry (DSC) measurements were carried out at different cooling/heating rates: 0.75, 1.5, 3, 4, 5 and 6 °C·min$^{-1}$. Exposure to moisture was minimized by drying the DSC pans in a vacuum oven at ~120 °C for ≥ 8 h, use of hermetic pans and sealing the samples inside of a dry Ar-filled glovebox (<0.1 ppm H$_2$O, < 1 ppm O$_2$) right before the measurement was conducted. Freezing temperatures (T$_{freeze}$) were obtained upon solidification of the samples, starting with a cooling step for the compositions that form room temperature (RT) molten salts. Alternatively, the compositions that are crystalline at RT were heated up to their melting temperature, followed by a cooling step.

**Molecular structures**
ACD/ChemSketch/3D Viewer (Freeware) 2021.1.0, version D25E41 was used to draw the structures in ball-and-stick style and converted into 3D. VESTA Ver. 3.4.8 was used to modify orientation and polyhedral style of the molecules.

**$^{27}$Al Nuclear Magnetic Resonance (NMR)**
Spectra were collected by using a Bruker AV400 NMR spectrometer at room temperature (298 K), BBFO broadband probe, HD electronics console. The quantitative acquisition was performed (128 scans, 90° excitation, 1s relaxation delay) with background suppression. The relative concentration determination of AlCl$_4^-$ and Al$_2$Cl$_7^-$ species were performed in MestReNova version 14.2.1-27684 (© 2021 Mestrelab Research S.L) with a qNMR plug-in.

**Electrode imaging and elemental mapping**

Plain carbon cloth working electrodes (WE) were collected from coin cells assembled for Coulombic efficiency measurements after an electrodeposition step was completed. Coin cells were disassembled inside an Ar-filled glovebox, cleaned with dimethyl carbonate (DMC), mounted on carbon tape, and examined by SEM (Zeiss Gemini 500) with an accelerating voltage of 5 kV, 20 mm aperture. The samples were transferred to the instrument using air-tight containers and loaded immediately after opening the containers to reduce exposure time to moisture. Elemental mapping was performed using Energy Dispersive X-Ray Spectroscopy-EDS/EDX with an accelerating voltage of 15 kV, 60 mm aperture.

The cross-section imaging of the Al electrode surface after cycling was conducted on Raith VELION Focused Ion Beam – Scanning Electron Microscopy (FIB-SEM) system with Au+ ion source. Before trenching, Pt was deposited onto the sample surface to protect the surface features. The cross-section images were taken using the SEM integrated into the system.

**Cell assembly and electrochemical measurements**

All cell assembly was conducted in a dry Ar-filled glovebox (<0.1 ppm $H_2O$, < 1 ppm $O_2$) and all electrochemical tests were performed at room temperature (~25°C). Al plating/stripping coulombic efficiency (CE) measurements were carried out in Al (CE) || carbon cloth (WE) and Al symmetric coin cells (CR2032) using a Neware battery tester. 0.9525 cm (3/8'') aluminum foil disks (25 µm thick, 99.45% metals basis, Alfa Aesar) were punch out and used as the anodes. 1.27 cm (½'') plain carbon cloth disks (356 µm thick, 1071 HCB, Fuel Cell Store) were used as a substrate for Al galvanostatic electrodeposition/dissolution. One layer of fiberglass fiber filter paper GF-D (Whatman) was placed between the WE and CE with a diameter of 1.905 cm or ¾'', and ~100 µL of electrolyte was poured before sealing. The amount of Al electrodeposited onto the carbon cloth in each step was fixed, followed by a stripping step. The percentage of Al stripped from the substrate electrode compared to the amount plated in the previous step represents the coulombic efficiency values reported in this work: $CE\,[\%] = (Stripping\ Capacity / Plating\ Capacity\ on\ substrate) \times 100$.

Electrochemical impedance spectroscopy (EIS) measurements were collected in Al symmetric cells at room temperature using a potentiostat BioLogic SP-200. The frequency range used was 7 MHz – 50 mHz, using a 10 mV as the perturbation voltage and 3 measurements were acquired per frequency. Al electrodes were used for ionic conductivity measurements. Equivalent circuit modeling was used to validate the analysis. The equivalent circuit used to analyze the frequency-dependent transport phenomena was $Z_{u+LE} + Q_{SL/SEI}/Z_{SL/SEI} + Q_{CT,\,Al-LE}/Z_{CT,\,Al-LE}$, where $Z_{u+LE}$ correspond to the uncompensated and liquid electrolyte impedance, $Q_{SL/SEI}$ and $Z_{SL/SEI}$ correspond to the constant phase element and ionic impedance of the solid layer/solid electrolyte interphase, respectively. $Q_{CT,\,Al-LE}$ and $Z_{CT,\,Al-LE}$ denote the constant phase element and impedance at the electrode / electrolyte interface (CT= charge transfer).

Full cell measurements were carried out in Al (CE) || graphite (WE) coin cells (CR2032) using a Neware battery tester. 0.9525 cm (3/8'') aluminum foil disks (aluminum foil disks (25 µm thick, 99.45% metals basis, Alfa Aesar) were punch out and used as the anodes. 1.27 cm (½'') plain carbon cloth disks (356 µm thick, 1071 HCB, Fuel Cell Store) were used as a substrate and current collector for the graphite cathodes. The graphite slurry consisted of 80 wt.% artificial graphite powder (MTI Corporation), 10 wt.% battery-grade carboxymethyl cellulose (CMC) binder, and 10 wt.% deionized water. After coating the carbon cloth substrates with the graphite slurry, the electrodes were dried in an oven at 90°C for ≥ 12 h before cell assembly. One layer of fiberglass fiber filter paper GF-D (Whatman) was placed between the anode and cathode, and ~100 µL of electrolyte was poured before sealing. The full cells were charged and discharged between the cut-off potentials of 2.3 and 0.2 V at a current density of ~330 mA·g$^{-1}$ (6 mA·cm$^{-2}$), except for the cells used for XRD and Raman characterization (1 mA·cm$^{-2}$).

Three-electrode cyclic voltammetry measurements were performed using a potentiostat/galvanostat Bio-Logic SP-200. The working electrode was a glassy carbon disk, the counter electrode an Al foil, and a 2 mm diameter Pt reference electrode (CH Instruments, Inc.). All three electrodes were placed in a t-shaped Swagelok-type cell with two layers of fiberglass fiber filter paper GF-D (Whatman) and ~200 µL of electrolyte. The scanning voltage range was set from -4 to 4 V (*versus* Pt.), and the scan rate was 20 mV·s$^{-1}$.

The ionic conductivity measurements as a function of temperature were conducted using a Novocontrol dielectric spectrometer with temperature control systems. The frequency range evaluated was from 10 MHz to 50 mHz, from 25 – 60 °C (± 0.5 °C).

**X-ray Photoelectron Spectroscopy Measurements**
A Surface Science Instruments SSX 100 was used for all XPS experiments. A custom-made airtight transfer holder was used to load the samples from a dry Ar-filled glovebox into the XPS instrument without air exposure. Survey scans used for the depth profiling used ion energy of 4 keV, a 10 mA current, a step size of 1 eV, a raster of 2 x 4 mm, 150 V pass energy, and a spot size of 400 µm. The high-resolution scans used a step size of 0.065 eV, resolution 2, while the rest of the parameters remained the same for the survey scans. All scans were quantified using Shirley backgrounds and sensitivity factors for O 1s, Al 2p, and Cl 2p in Casa XPS software. Core scans used a pass energy of 150 V and were calibrated using C-C bond energy at 285 eV. The following time steps and sequence were used to obtain surface chemistry measurements of the Al anode at the surface and near the bulk; 0, 2, 5 min to remove adventitious carbon, 3 measurements every 10 min of Ar-sputtering, a high-resolution scan, 3 measurements every 30 min of Ar-sputtering, a high-resolution scan, and 5 measurements every 60 min of Ar-sputtering, for a total of of ~13 h.

**Structural changes of graphite cathodes upon intercalation**
Graphite cathodes were collected after charging up to 1.9, 2.0, 2.1, 2.2 and 2.3 V from Al || graphite cells utilizing either AlCl$_3$: TriMAHCl (1.5:1) or AlCl$_3$: TriEAHCl (1.8:1) electrolytes. Cathodes were rinsed with anhydrous methanol, dried, and kept under an inert atmosphere prior to analysis.

X-ray diffraction measurements were carried out using a Bruker D8 powder diffractometer (Cu kα 1.54 Å radiation, step size 0.0194577°) between 20-35 2-theta degrees at 40 kV, 25 mA, a divergent beam slit of 1.0 mm, and a detector slit of 9 mm.

Raman spectra were collected using a WITec-Alpha 300R Confocal Raman Microscope. A 532 nm green laser was used at 2 mW, 1200 l·mm$^{-1}$ grating, spectral center at 1500 cm$^{-1}$, 15 accumulations, and 30 s integration time.

**Intercalation Species Determination *via* X-ray Absorption Spectroscopy Measurements**
Cathodes were collected in the charged state (2.3 V) from Al || graphite cells utilizing either AlCl$_3$: TriMAHCl (1.5:1) or AlCl$_3$: EMIM]Cl (1.8:1) electrolytes. Cathodes were rinsed with anhydrous methanol, dried, and kept under an inert atmosphere prior to analysis. Al K-edge X-ray Absorption spectroscopy measurements were collected at the National Institute of Standards and Technology Partner Beamline 7-ID-1 (SST-1) at the National Synchrotron Light Source II at Brookhaven National Laboratory. Measurements were recorded under 10$^{-7}$ Pa vacuum. In the pre-edge range (-200 to -20 eV below the edge), the incident beam energy was scanned using 5 eV steps. Across the edge (-20 – 30 eV), a 0.3 eV step size was used for enhanced resolution, and in the post edge region (20 – 800 eV) a step of 0.05 Å$^{-1}$ was used. A 1s acquisition time was used at each data point. Data collected in total electron yield (TEY) detection mode was used for the XAS analysis and fluctuations in the incident beam were accounted for via normalization using an Au coated mirror located upstream of the sample. The energy was calibrated by setting the maximum of the first derivative of an Al foil spectra to 1559 eV. XAS spectra were aligned, averaged, and normalized using Athena. (*1*) The built-in AUTOBK algorithm was used to minimize background below R$_{bkg}$ = 1.2 Å. Normalized spectra were fit utilizing Artemis using theoretical models for AlCl$_4^-$ and Al$_2$Cl$_7^-$ created with FEFF6. (*1, 2*) based on reported crystal structures of chloroaluminate salts. (*3, 4*) A k-range of 2.5 –7.5 Å$^{-1}$ and Hanning window (dk = 1) were used as Fourier transform parameters, and fitting was performed using

k$^2$ weighting. The R-space window was defined as 1.2 – 4.0 Å. The S$_0^2$ parameter was determined to be 0.88 from fitting an Al metal foil standard, and this term was applied to the fits to account for intrinsic losses in the electron propagation and scattering processes. (*5*)

The probe depth of the total electron yield (TEY) XAS measurements must also be considered when interpreting the experimental results.(*6*) During the X-ray absorption process a primary core level electron is ejected, and the resulting core-excited atom can undergo relaxation either via emission of a fluorescent photon or via emission of an Auger electron. (*7–11*) Inelastic scattering of primary Auger electrons also occur, resulting in emission of secondary electrons. In the TEY detection mode, the total number of emitted electrons (primary, Auger, secondary) from the sample are measured. The TEY probe depth increases with the core level energy being probed, with experimental studies indicating a maximum probe depth of 14.1 nm at 0.929 keV (Cu-L edge). At the Al-K-edge (1.559 keV), the attenuation length of incident X-rays will be lower and the kinetic energy and inelastic scattering range of the Auger electrons will be higher, and consequently the probe depth of the TEY measurement in this experiment is anticipated to be greater than 14 nm.(*9*) While the staging mechanism of AlCl$_4^-$ intercalation in graphite is not completely understood,(*11–13*) average interlayer spacing from XRD has been reported between 4.55 Å (*14*) and 5.70 Å (*7*) , and a probe depth of 15 nm would correspond to 25 – 30 graphene layers in the intercalated graphite electrode.